# The Large Smooth Patch on Comet 9P/Tempel 1: Remnant of a Recent Past Event

J. L. Rizos[1,2], T. L. Farnham[2], J. M. Sunshine[2], J. Kloos[2], J. L. Ortiz[1]

1 - Instituto de Astrofísica de Andalucía, Glorieta de la Astronomía S/N, Granada 18008, Spain.
2 - University of Maryland, Department of Astronomy, 4296 Stadium Drive, College Park, MD 20742-2421, USA.

**Abstract**

We present a comprehensive reassessment of the region containing the large smooth patch on comet 9P/Tempel 1, leveraging data from the Deep Impact and Stardust-NExT missions, an updated stereophotoclinometry-based shape model, and numerical simulations. The study seeks to understand the nature, the triggering mechanism, and the chronology of this distinctive feature. Morphological and spectral analysis reveals that the smooth patch has a thickness of approximately 25 meters, a notable lobate U-shape, and a spectral composition indistinguishable from the surrounding terrain, which favors an endogenous origin. Gravitational flow simulations demonstrate that a single event could have formed the large smooth patch, the secondary smooth units observed on other faces of the comet, and the topographic terrace features adjacent to the northern smooth unit. We estimate this event occurred between 600 and 1,200 years ago, a temporal window that notably coincides with a period of abrupt orbital changes caused by multiple close encounters with Jupiter. We propose that these encounters may have played a role in triggering a mass flow. Although with the underlying mechanism still unresolved, these results shed new light on the geology of cometary nuclei and on the role of external dynamical processes in shaping their surfaces.

## 1. Introduction

Comets are among the most pristine bodies in the Solar System, formed in the outermost regions of the solar nebula. Unlike main-belt asteroids or terrestrial planets, they contain volatiles—most notably water ice—that remained solid in the cold outer region of the Solar System (Weidenschilling 2004).

From a dynamical perspective, comets are commonly divided into two broad categories according to their source reservoirs: isotropic comets, thought to originate in the Oort Cloud and characterized by nearly random orbital inclinations and periods ranging from hundreds to millions of years; and ecliptic comets, which arise from trans-Neptunian populations in the Kuiper Belt and Scattered Disk (Rickman 2004).

Among ecliptic comets, a notable subgroup consists of the Jupiter-family comets (JFCs), conventionally defined by Tisserand parameter with respect to Jupiter in the range $2 < T_J < 3$. They typically have aphelia near Jupiter's orbit and relatively low inclinations with respect to the ecliptic. Most JFCs are thought to originate from the Scattered Disk population beyond Neptune (Duncan 2008). Their present-day orbits are strongly shaped by repeated gravitational interactions with Jupiter, which both control their dynamical evolution and give the class its name. Observational and dynamical studies suggest that many JFC nuclei are collisionally processed bodies from the trans-Neptunian region (Stern 1995; Farinella & Davis 1996), implying that they may not fully preserve the pristine state. Nonetheless, as descendants of trans-Neptunian populations, JFCs provide a valuable dataset for probing their progenitors and the broader evolutionary processes of the Solar System. Thanks to their short orbital periods, they frequently approach the Sun, offering favorable conditions for observation. To date, resolved surface images have been obtained for six comets visited by spacecraft: 1P/Halley (1986, Giotto), 19P/Borrelly (Deep Space 1, 2001), 81P/Wild 2 (Stardust, 2004), 9P/Tempel 1 (Deep Impact + Stardust-NExT, 2005 & 2011), 103P/Hartley 2 (EPOXI, 2010) and 67P/Churyumov–Gerasimenko (Rosetta, 2014) (Keller et al., 1986; Nelson et al., 2004; Tsou et al., 2004; A'Hearn et al., 2005a; Veverka et al., 2013; A'Hearn et al., 2011; Rickman et al., 2015).

Comet 9P/Tempel 1, with an estimated diameter of approximately 6 km, is classified as a JFC ($q$ = 1.54 au, $Q$ = 4.75 au, $i$ = 10.47°). Discovered in 1867, it became the target of two space missions: Deep Impact (DI) in 2005 (A'Hearn et al., 2005a) and Stardust/NExT (SDN) in 2011 (Veverka et al., 2013). As the only comet observed during two separate apparitions, 9P/Tempel 1 offers a unique opportunity to document surface changes and study its physical properties and evolution over a full orbital period of 5.56 years.

The DI mission successfully directed a 366 kg impactor spacecraft to collide with 9P/Tempel 1's surface at a velocity of 10.2 km/s (A'Hearn et al., 2005b). The spacecraft was equipped with four scientific instruments: three digital cameras—the High-Resolution Instrument (HRI), Medium Resolution Instrument (MRI), and Impactor Targeting Sensor (ITS)—and an infrared (IR) imaging spectrometer that shared the HRI telescope as its fore optics (Klaasen et al., 2008). In contrast, the SDN spacecraft, during its extended mission to 9P/Tempel 1, carried a payload consisting of a visible camera (NavCam) and two dust detection instruments (Klaasen et al., 2013). Combined, the images captured by these missions covered approximately 70% of 9P/Tempel 1's surface (Thomas et al., 2013), although subject to differences in spatial resolution, dynamic range, and viewing geometry. These images reveal a morphologically diverse body characterized by rough, pitted terrain along with distinct smooth regions.

These smooth units, identified by Thomas et al., (2007)—who suggested a possible interconnection and hypothesized their presence in the unobserved regions of the comet—are distributed across multiple areas of the nucleus. The most prominent of these, and the primary focus of this study, is a large smooth deposit (hereafter referred to as the "*large smooth patch*"; see Figs. 1c, 2 and A1), over 2.5 km in length, located near the comet's south pole. This deposit is surrounded by rugged terrain and lies within a recess bounded by a pronounced topographic break "*cliff*". Here, we use the term cliff in a morphological sense, referring to a marked change in slope rather than to a specific mechanical classification. Throughout this work, we designate the area between the edge of the smooth patch and this topographic break as the "*cliff region*".

In addition, two secondary smooth units are present. One is situated in the equatorial region, near the large depression where Sunshine et al., (2006) detected three localized exposures of water ice.

We refer to this as the "*equatorial smooth unit*" (see Fig. 1a and A2). The other is located on the northern face of the nucleus, roughly opposite the large smooth patch. This "*northern smooth unit*", revealed by images from the SDN mission (see Fig. 1c and A3), remains only partially characterized due to limited image coverage. Adjacent to this northern unit is a topographic terrace featuring up to five distinct ridges or steps along its length (Thomas et al., 2013), which appears to be the result of a mass-wasting process.

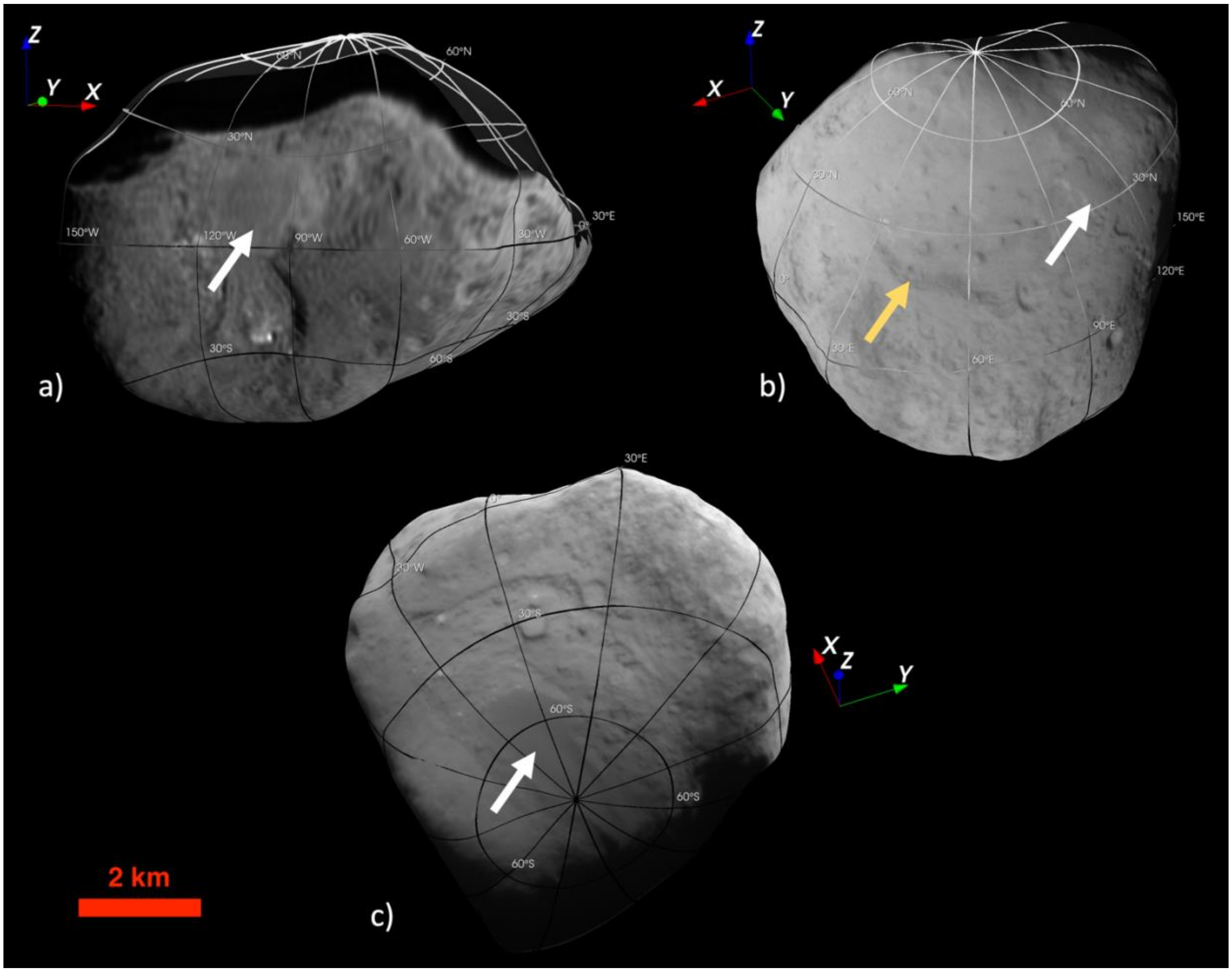

**Fig. 1. Global views of comet 9P/Tempel 1 from different perspectives, providing overall context for the regions discussed in this work. In all panels, the Z-axis indicates the north direction. (a) Equatorial region observed in the deconvolved HRI image 9000909; the arrow marks the secondary smooth unit, hereafter referred to as the *equatorial smooth unit*. (b) Northern hemisphere as seen in Stardust-NExT image 30039; the white arrow indicates a region where a smooth unit is visible, hereafter referred to as the *northern smooth unit*. The orange arrow highlights a system of terraces comprising up to five distinct steps. (c) Southern region observed in Stardust-NExT image 30035; the arrow marks the large smooth patch.**

Although the best-imaged comet to date, 67P/Churyumov–Gerasimenko, does exhibit kilometre-scale smooth terrains (e.g. the Imhotep region), these units are observed to evolve on seasonal timescales and are generally interpreted as thin, mobile deposits produced by dust fallback and granular transport (Keller at al., 2017). In contrast, the large smooth patch on 9P/Tempel 1 represents a substantially thicker deposit, bounded by rugged, pitted terrain and a cliff, suggesting a distinct geomorphic setting and, consequently, a different formation mechanism. On the other hand, observations of 19P/Borrelly pointed to a smooth region that appears to be an isolated kilometer scale mesa (Britt et al., 2004). However, as it was noted by Thomas et al., (2007), in the lower resolution of Borrelly images it cannot be definitively confirmed. For this reason, the large smooth patch on 9P/Tempel 1 represents a unique geomorphological feature among well-resolved JFCs to date.

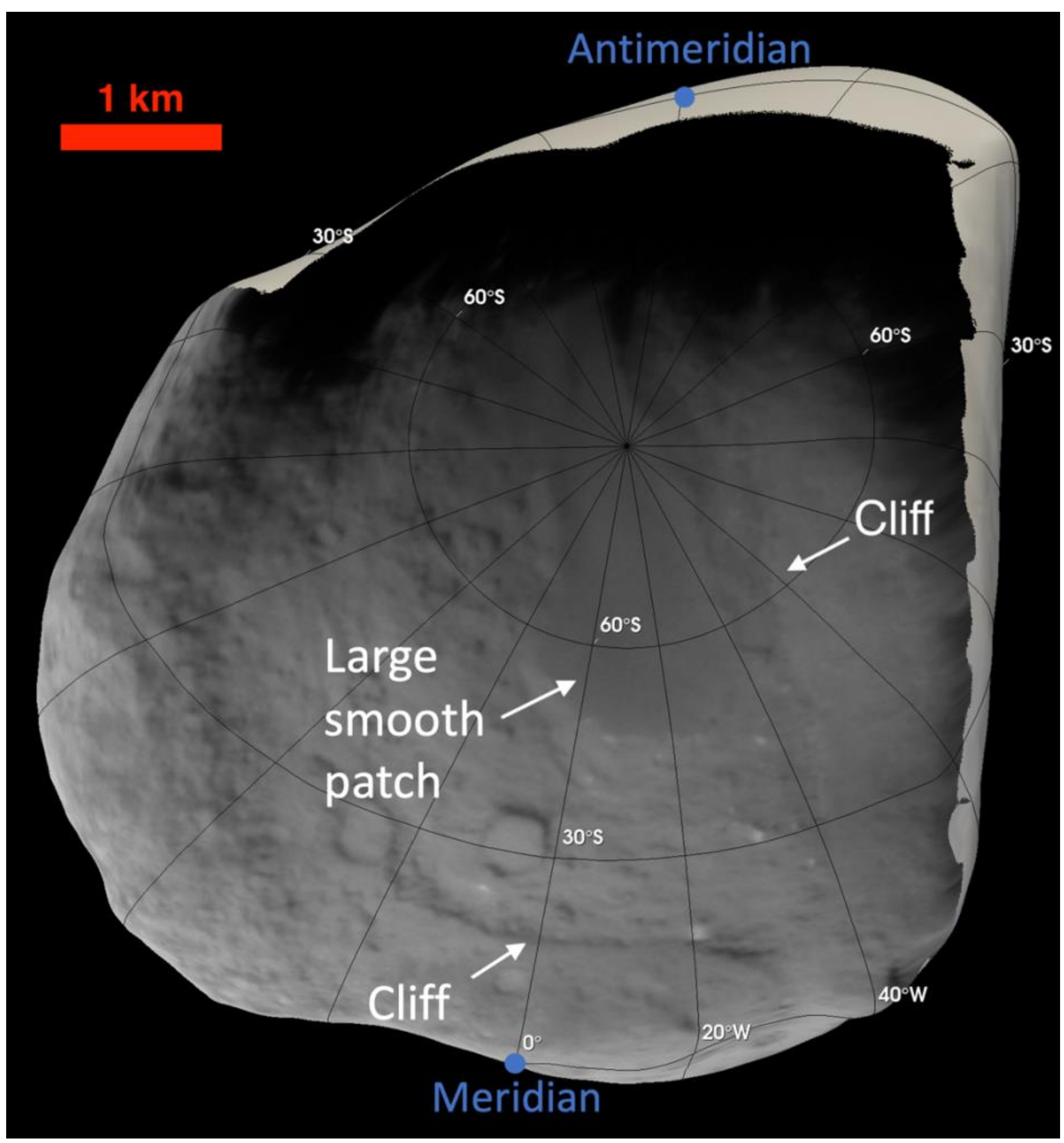

**Fig. 2. Same region as shown in Fig. 1c, but from a different perspective, fully centered on the southern hemisphere. The low albedo smooth region in the center is the large smooth patch, surrounded by rugged terrain and a cliff. The blue dots mark the meridian (longitude 0°) and the antimeridian (longitude 180°), used as reference points to orient the reader throughout the following figures.**

The origin of 9P/Tempel 1's large smooth terrain and its surrounding features remains elusive, and understanding this distinctive region may hold the key to advancing our knowledge of cometary evolution. Since its discovery, several hypotheses have been proposed. Belton et al., (2007) suggested that the smooth terrains represent primitive layers exhumed by sublimation; Bar-Nun et al., (2008) argued that these terrains result from the deposition of ice grains ejected in a collimated outburst of gas; and Belton and Melosh (2009) proposed a complex mechanism involving subsurface fluidization. However, none of these models discuss the specific geomorphological configuration described here, particularly the relationship between the smooth patch and the adjacent cliff. To date, the only hypothesis that simultaneously addresses the geometry of the smooth patch, its downslope orientation, and its relationship with the adjacent scarp is a material flow interpretation (Thomas et al., 2007). By measuring changes in features in images between DI and SDN, Thomas et al., (2013) estimated that this smooth patch eroded at a rate of 2 – 4 × $10^5$ $m^3$/orbit, and assuming a constant rate of erosion, suggested that the current state would have been reached in at least 150 years, placing a lower age limit on this structure. Supporting this idea, Goguen et al., (2008) calculated that with a flow velocity below the surface escape velocity (1.3 m/s), the material's kinematic viscosity must exceed $2.5\times10^{-3}$ $m^2$/s, and the flow must exhibit a high Reynolds number (~$10^4$), indicative of turbulent behavior. This aligns with Basilevsky and Keller (2007), who considered eruptions from an active interior unlikely and instead proposed a sublimation-driven collapse of steep slopes, resulting in an avalanche of material into low-lying areas.

The aim of the study presented here is to contribute to the ongoing discussion by reassessing previous data and leveraging an updated stereophotoclinometry-based shape model (Ernst et al., 2019), which offers a substantial improvement in spatial resolution—reducing the ground sample distance from ~80 m in the previous model (Farnham & Thomas 2013) to as little as 8 m. To this end, we analyze the large smooth patch using images from DI HRI, MRI, and ITS, and SDN NavCam, complemented by IR spectral data from the DI mission. Section 2 describes the data and methods employed in this study. We first present a spectral analysis of the DI data to assess potential compositional differences. We subsequently quantify the thickness and spatial extent of the large smooth patch and its surroundings using the new shape model, complemented by anaglyphs to visualize its three-dimensional morphology. Next, we perform a comparative analysis of DI and SDN images to identify and classify morphological changes within the region. We then examine the solar irradiation over a complete orbital period, followed by a numerical simulation of the hypothesized gravitational flow on the surface. Finally, we discuss the various hypotheses compatible with our results and other recently published studies.

## 2. Data and methods

### 2.1 Datasets

*DI images*: Images acquired by the DI spacecraft, stem from the HRI, MRI, and ITS instruments. The HRI features a Cassegrain telescope with a focal length of 10.5 m and a 30 cm aperture. It serves a dual purpose, directing light to a filtered CCD camera (1024 x 1024 pixels) and a long-slit imaging spectrometer via a dichroic beamsplitter. This beamsplitter reflects visible light (0.34 to 1.05 m wavelength) to the CCD while transmitting IR light (1.05 to > 4.8 m). The visible light is filtered through one of nine color filters centered at 350, 450, 550, 650, 750, 850, and 950 nm, respectively. Note that HRI was discovered to be out of focus post-launch (Klaasen et al., 2008). However, there is a set of deconvolved images (Lindler et al., 2007) that enhance the resolution, although they are not useful for photometric measurements due to artifacts produced by the deconvolution algorithm. MRI and ITS, both 1024 x 1024 pixels, are similar instruments based on a 2.1 m Cassegrain telescope with a 12 cm aperture. MRI natively has five times lower spatial resolution than HRI and features a nine-position filter wheel centered at 309, 345, 387, 514, 526, 650, 700, 750, and 950 nm. ITS, in contrast, is a monochromatic imager. The images utilized herein are accessible via the Planetary Data System (PDS[1], McLaughlin et al. 2014a, 2014b, 2014c). Detailed information on their calibration and additional technical specifications can also be found both in the PDS repository and in Klaasen et al., (2008).

*SDN images*: The images from the SDN spacecraft were acquired using the NavCam instrument. This camera, with a 0.2 m focal length, was connected to a 1024 x 1024-pixel CCD. Despite the camera's original design to use various color filters, issues with the filter wheel during flight resulted in images taken only with the broadband optical-navigation spectral filter (475 to 925 nm; Klaasen et al., 2013). As with the DI images, the calibrated images analyzed in this study were retrieved from the PDS[2] (Veverka et al., 2011), with detailed calibration and technical information available both in the PDS repository and in Klaasen et al., (2013).

*Shape model:* We utilize a stereophotoclinometry model of 9P/Tempel 1 as described by Ernest et al., (2019), which merges DI and SDN data. In contrast to the previous shape model (Farnham & Thomas 2013), this new model enables the incorporation of local, high-level details like those observed in the images or spectra. This updated shape model, consisting of 3,145,734 facets with a ground sample distance of up to 8 meters, has been refined using 318 HRI, 379 MRI, 83 ITS, and 72 NavCam images, resulting in an enhancement in resolution compared to the previous version of 32,040 facets. The HRI images used correspond to the original out-of-focus products rather than the deconvolved versions.

### *2.2 Methods*

*Gravitational flow simulation*: Thomas et al., (2013) interpreted the large smooth patch observed in DI images as deposits resulting from flows. To examine this hypothesis, we first computed the gravitational potential energy across the surface of 9P/Tempel 1. To achieve this, we subdivided the shape model into triangular pyramids with their apexes positioned at the origin coordinates, following the approach outlined by Chanut et al., (2015).

The simulation begins by placing a seed at a point on the surface, treating that point as the source from which a flow originates. The gravitational potential of this source is then compared with those of its neighboring facets. If a neighboring facet has a lower gravitational potential, the flow advances to that facet in the next step. This process is iterated until no new facets can be occupied.

For the simulation, we assumed a mean density of 400 kg/m$^3$ (Richardson et al., 2007), an average gravity ($g_r$) of 2.7 x 10$^{-4}$ m/s$^2$ (Thomas et al., 2007) and a homogeneous distribution of mass. The rotational period of 9P/Tempel 1, approximately 40 hours (Belton et al., 2011), induces negligible additional acceleration on the surface gravity field (Thomas et al., 2013), and was therefore disregarded. To reduce the computational cost, the shape model was downsampled by a factor of 100, resulting in a mesh of 31,456 facets.

Note that this model is purely kinematic and evaluates only the gravitational compatibility between potential source regions and topographic lows. It does not compute viscous stresses, basal shear forces, or material strength, and therefore does not attempt to reproduce the mechanical details of erosive or rheological flow. In addition, it does not account for possible internal heterogeneities, long-term mass redistribution, or changes in spin state, as the relevant physical and mechanical parameters for 9P/Tempel 1 remain insufficiently constrained.

[1] https://pds-smallbodies.astro.umd.edu/data_sb/missions/deepimpact/index.shtml

[2] https://pdssbn.astro.umd.edu/holdings/sdu-c_cal-navcam-3-next-tempel1-v1.0/dataset.shtml

To comprehensively explore all solutions, the entire surface is divided into a 200 × 200 meters grid, and seeds are systematically placed at each grid point. Rather than employing units of gravitational potential energy, which can be challenging to intuitively relate to height in a shape as irregular as 9P/Tempel 1, we opt to transform the gravitational potential ($W_i$) for the $i$-th facet into dynamic height, denoted as $D_i = \frac{W_i - W_0}{g_r}$, where $W_0$ represents the lowest gravitational potential energy observed on the surface (Thomas et al., 1993).

*Solar irradiance*: We compute the orbit-integrated absorbed solar energy distribution across the surface of comet 9P/Tempel 1 using its shape model and SPICE ephemerides, assuming the current present-day orbital configuration and spin-pole obliquity.

For each triangular facet of the shape model, we determine the outward unit normal vector in the body-fixed reference frame. At each epoch, the Sun–comet vector is obtained from SPICE and expressed in the inertial J2000 frame. The facet normals are rotated into the same inertial frame to ensure a consistent geometric calculation. The solar incidence angle, $i$, is then computed from the dot product between the rotated facet normal and the Sun direction vector. Only illuminated facets (i.e., those with $\cos(i) > 0$) contribute to the energy budget. The heliocentric distance is calculated at each epoch and expressed in astronomical units.

The instantaneous absorbed irradiance (W/m$^2$) for the $j$-th facet is calculated as:

$$E_j = \frac{(1 - A)F_0 cos(i_j)}{d^2} \qquad (1)$$

where $A$ is the surface albedo (0.04), $F_0$ is the solar flux at 1 AU (1361 W/m$^2$), and $d$ is the heliocentric distance in AU. We verified that the contribution from self-heating (facet-to-facet thermal radiation) is negligible and therefore it is not included in the calculations.

The calculation is performed over one complete orbital period, subdivided into 12,200-time steps, corresponding to a temporal resolution of approximately 4 hours. Given the ~40-hour rotation period of Tempel 1, this sampling provides roughly ten evaluations per rotation. The absorbed irradiance is integrated over time by multiplying each instantaneous facet value by the corresponding time interval, yielding the orbit-integrated absorbed solar energy per unit area (J/m$^2$).

*3D morphologic measurements*: To capture three-dimensional data of this area, we employ photogrammetry techniques in combination with the shape model. In particular, we use the shadows cast by the central depressions within the large smooth patch to characterize the local surface curvature. For a convex surface, we can establish a trigonometric relationship between the length of the shadows ($L$), the incidence angle at the shadow's edge ($i$), and the relative elevation of the endpoint ($h$) as $h \sim L\ tan(90 - i)$ (see Fig. 3a). We refer to this approach as the "shadow–length method". Figure 4a illustrates an example of its application to our case.

In addition, we use images acquired at different emission angles, in which the edges of the large smooth patch are clearly resolved. The projected length of an edge as seen from the camera can be expressed as $S_{proj} = S\ cos(e)$, where $e$ represents the emission angle. Using the trigonometric relationship of $h = S\ sin(\alpha)$ (see Fig. 3b), and assuming non-extreme geometries in which $e \approx \alpha$ and viewing conditions are not near the limb (i.e. $e \not\approx 90°$), the thickness can be approximated as $h \approx S_{proj}\ tan(e)$. We refer to this approach as the "projected-length approximation". Figure 4b shows an example of its application to our case.

To assess the validity of these assumptions and approximations, we identified control points that are sufficiently elevated to be well defined in the shape model, measured their heights directly from the model, and verified that both methods yield consistent results.

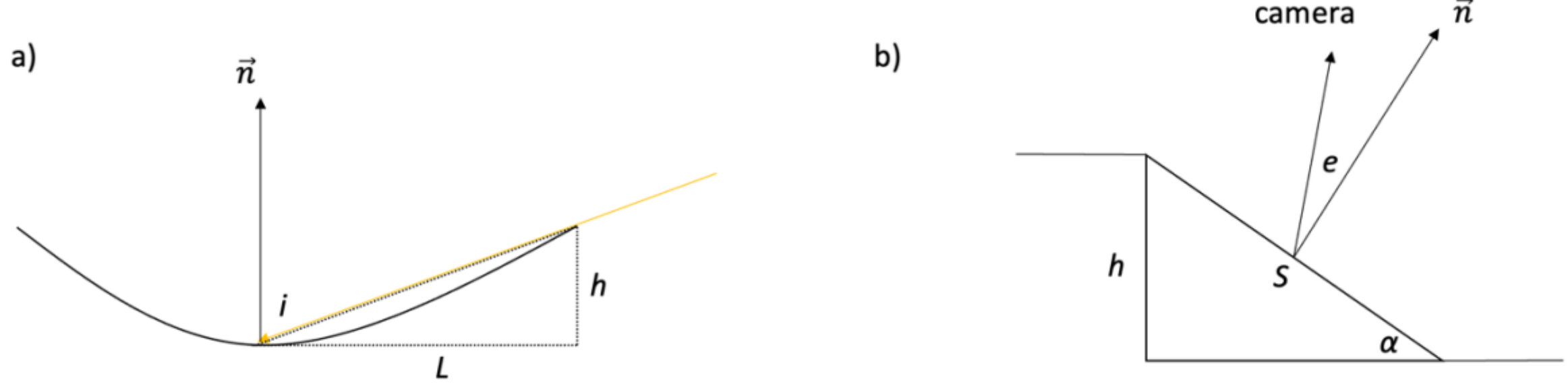

**Fig. 3. a) Diagram showing the smooth patch surface as seen by NavCam where we establish a relationship between the relative elevation of the endpoint (*h*), the incidence angle (*i*), and the length of the shadow (*L*). b) Diagram of the smooth patch edge seen by ITS. The emission angle (*e*) and the edge length (*S*) can be related to get the order of magnitude of the thickness (*h*).**

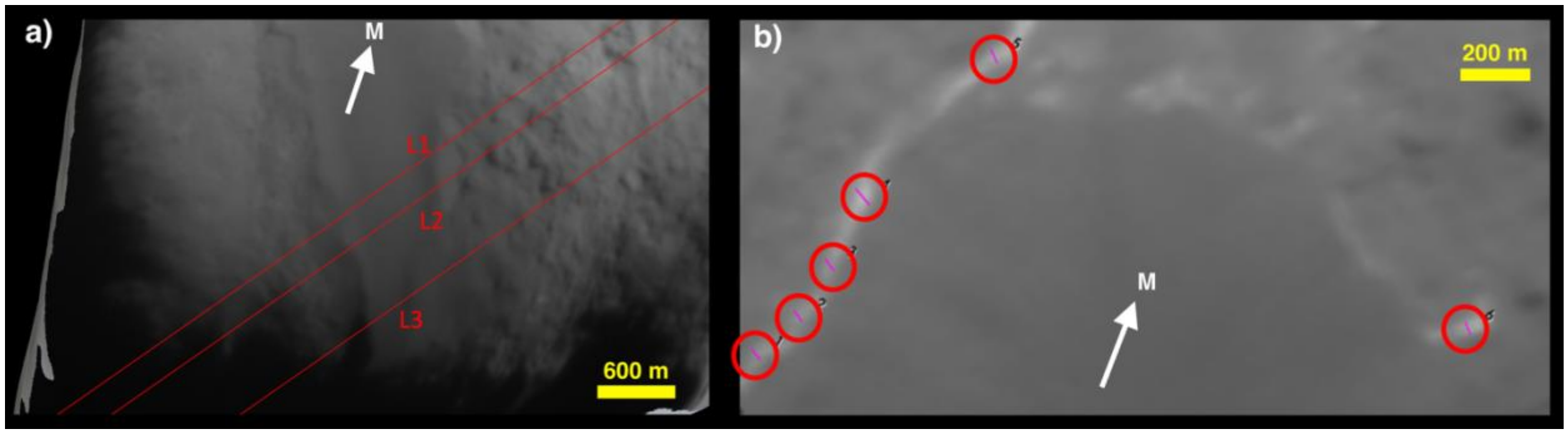

**Fig. 4. a) Example of an SDN image in which shadows caused by the convex shape of the smooth patch are visible. The red lines are aligned with the direction of the Sun and were used to measure the length of the shadows. b) Example of an ITS image showing the edges of the large smooth patch as seen from the camera. The red circles represent multiple measurements taken along the entire edge. In both figures, the white arrows indicate the meridian (see Fig. 2 for reference).**

In addition to the methods described above, we also generated anaglyphs (Fig. 7), which consist of pairs of stereo images acquired from slightly different viewing geometries. We selected image pairs with comparable spatial resolution but distinct viewing angles. The RGB channels were adjusted to produce red–cyan composite images. The resulting images were then rotated to ensure horizontal alignment for the observer, and a controlled lateral offset was applied until the desired three-dimensional effect was achieved when viewed with red–cyan glasses.

## 3. Results

### 3.1 Spectral data

Using the HRI color filters, we measured the visible spectra from four distinct regions of 9P/Tempel 1, as indicated in Fig. 5. HRI was selected instead of MRI because its superior spatial resolution enables a more accurate characterization of localized surface variations. To facilitate comparison, we normalized the values at 650 nm. The error is the standard deviation of each square (Fig. 5a). This process was repeated for the IR spectra, which were normalized at 2 microns (Fig. 5d). All spectra were obtained at a phase angle of 62.9°. Region 1 specifically targets an area withing the large smooth patch. Interestingly, no significant spectral disparities were observed in any case, indicating that the composition of the four regions exhibits no measurable differences. With the exception of localized ice exposures (Sunshine et al., 2006), 9P/Tempel 1 exhibits a compositionally uniform surface.

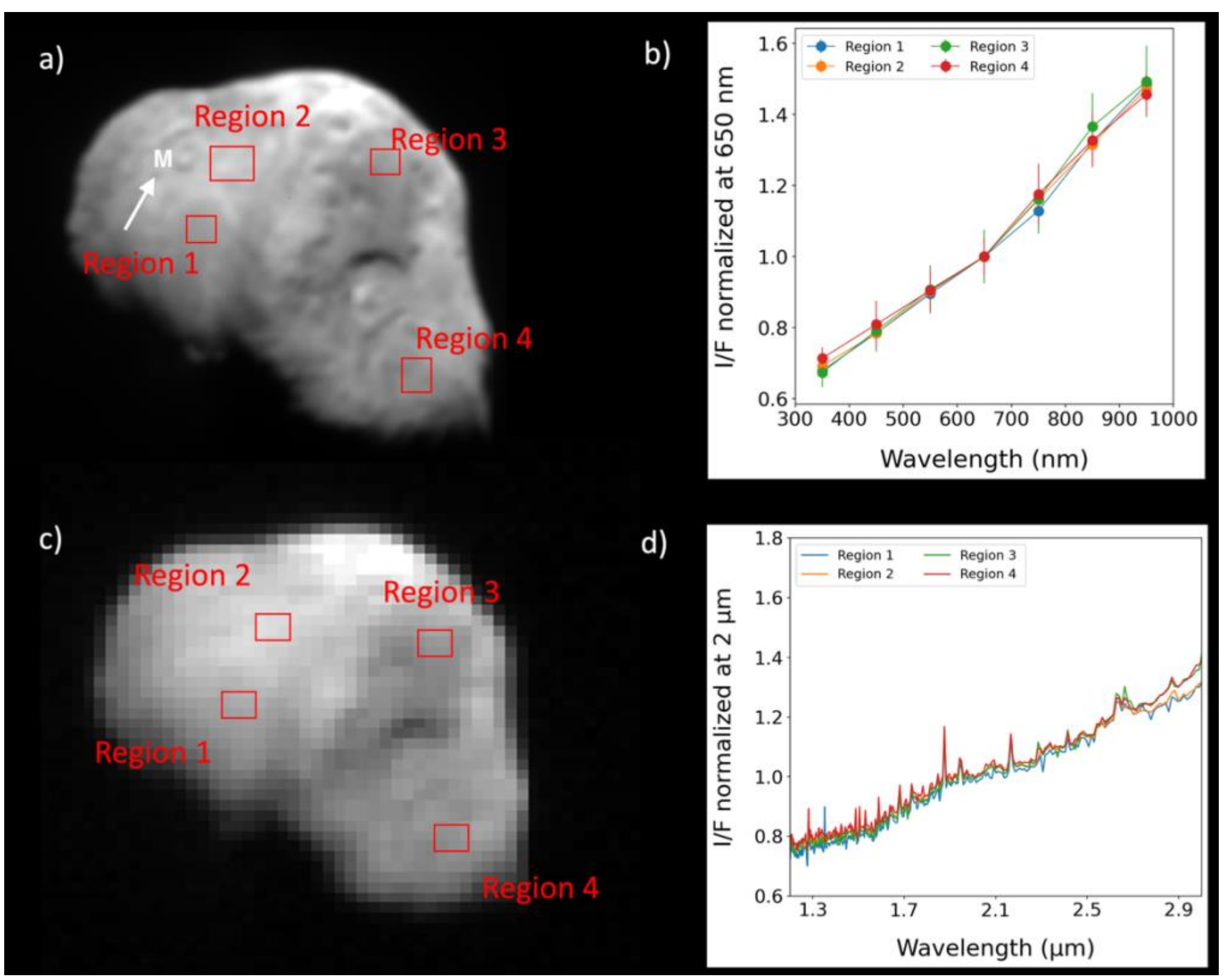

**Fig. 5. Visible and infrared observations of comet 9P/Tempel 1 from four selected regions (red squares). Panel (a) shows the HRI visible image used to extract spectra from the marked areas, with the corresponding reflectance spectra obtained in panel (b). Panel (c) presents a reconstructed infrared image with the same regions indicated, and panel (d) shows their corresponding IR spectra. Region 1 corresponds to the large smooth patch in both images. No significant spectral variability is observed among the selected regions. The white arrow indicates the meridian (see Fig. 2 for reference).**

## 3.2 Spatial Measurements from Image Data

First, we conducted measurements by projecting the images onto the updated shape model. Since the shape model captures the elevation of the cliff that borders the large smooth patch, we selected 16 points on both sides and obtained the relative height of pairs a-b, one in the inner lower region and the other in the outer higher region, as shown in Fig. 6a. This analysis was performed using the Small Bodies Mapping Tool software (SBMT[3], Ernst et al., 2018). Because this body is not a spheroid, the height cannot simply be derived from the radius differences of each pair. Instead, we projected the shorter vector onto the longer one, obtaining an average height for the cliff of 0.054 ± 0.017 km.

Additionally, using SBMT we delineated polygonal boundaries for both the smooth patch and the surrounding cliff-bounded region (Fig. 6b). The smooth patch covers an area of 1.720 ± 0.010 km², while the cliff-bounded region, defined as the light red area (excluding the dark red smooth patch) in Fig. 6b, covers an area of 7.140 ± 0.010 km². It is important to note that the boundaries defining the cliff edge and the large smooth patch are not always precisely discernible, and shadows in certain regions further complicate this task. Thus, these area measurements are approximate and should be interpreted as order-of-magnitude estimates for the extent of these regions.

To investigate the morphology of the large smooth patch and quantify the surface curvature, we utilized the shadow-length method (refer to Section 2.2). For that, three parallel lines towards the Sun are drawn on the 9P/Tempel 1 surface (Fig. 4a). This yields elevation differences of 0.021 ± 0.010, 0.018 ± 0.010, and 0.009 ± 0.010 km for L1, L2, and L3, respectively, indicating a systematic increase in elevation away from the central region of the smooth patch. The 0.01 km elevation uncertainty is estimated from the validation exercise, where the method is applied to shadow-based control points (Fig. 6a, control points 17–19) and compared with direct elevation measurements from the shape model, yielding agreement at the ~10 m level. Fig. A1 shows this region together with illustrative cross-sectional profiles to help the reader visualize the geometry and better interpret the results discussed in this section. This outcome reveals that the central region of the smooth patch is lower than its sides, forming a distinct U-shape.

We also created anaglyphs (Fig. 7) using three pairs of images from the ITS, HRI, MRI, and SDN instruments, selected based on suitable changes in viewing geometry. Fig. 7a is composed of the 5070405_9000639 ITS image (cyan) and the deconvolved 5070405_9000904 HRI image (red). Fig. 7b is composed of 30034 (cyan) and 30035 (red) SDN images. Post-impact Fig. 7c is composed of MRI images, 5070405_9000999 (cyan) and 5070405_9001012 (red). The three-dimensional view provided by these anaglyphs confirm the U-shaped morphology.

[3] https://sbmt.jhuapl.edu/

To measure the thickness of the smooth patch, we applied the projected-length approximation described in Section 2.2 to ITS images (Figs. 3b and 4b), measuring the emission angle and the projected edge length in several sections. This yields an estimated thickness of 0.025 ± 0.005 km, with the uncertainty representing the standard deviation of the individual thickness measurements. This projected-length approximation is applicable because the geometry does not correspond to a near-limb case, as the emission angle in this region is approximately 43°. To validate this approximation, we applied it to well-defined border features of the pit crater (control-point pairs 17 and 18; Fig. 6a), which can be measured directly on the shape model. The resulting values confirm that this approximation is valid for our case. This value is consistent with the observations of Thomas et al., (2007), who previously indicated that it should be at least 0.020 km thick, although there is no record of exact measurements or method.

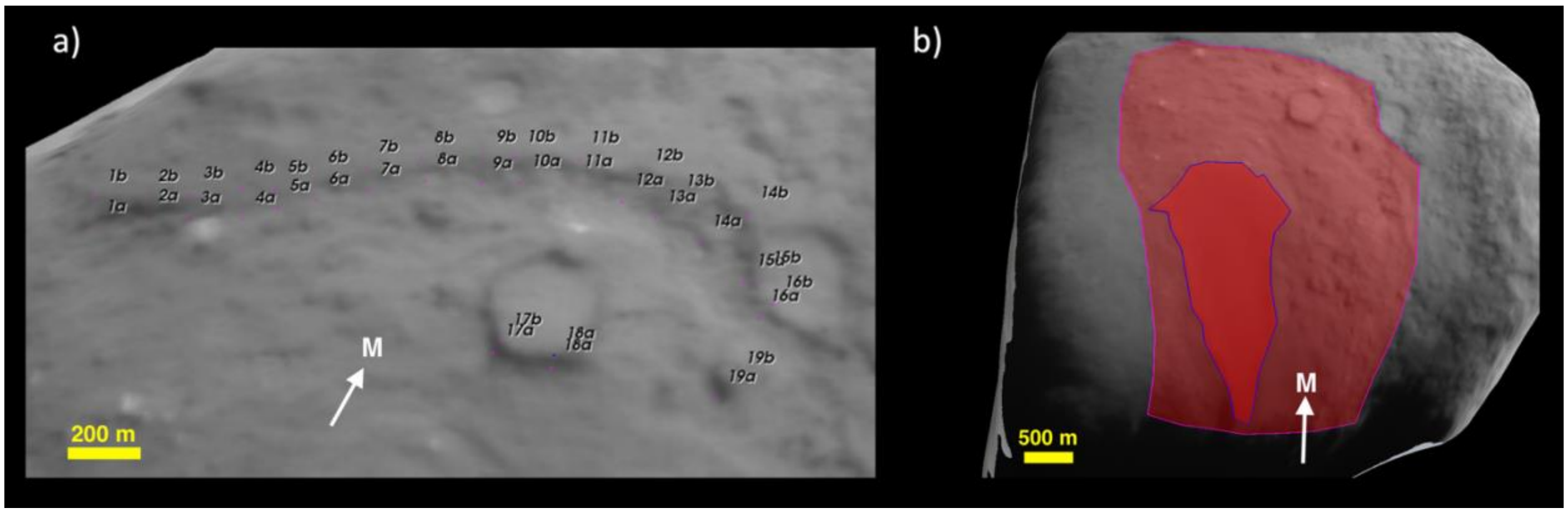

**Fig. 6. Example of measurements of the cliff region and the smooth patch area using the SBMT software. a) Pairs of points used to measure cliff elevations using the shape model. b) Area measurements after projecting images onto the shape model. The smooth patch is indicated by the dark red region. The surrounding light red region delineates the cliff-bounded region. The cliff area for which we report area and mass corresponds to the light red region, excluding the dark red smooth patch. In both figures, the white arrows indicate the meridian (see Fig. 2 for reference).**

Using this mean effective thickness and the projected area of the smooth patch, we estimate a volume of 0.043 ± 0.009 km³. Assuming a uniform density of 400 kg m⁻³ (Richardson et al. 2007), this corresponds to a mass of (1.72 ± 0.34) × $10^{10}$ kg. Although the surface of the smooth patch exhibits a concave, U-shaped morphology, we neglect this curvature in the volume and mass estimates.

On the other hand, the volume of the surrounding cliff-bounded region (light red region in Fig. 6b), is estimated to be 0.386 ± 0.121 km³, which represents only 0.4% of the total volume of Tempel 1. Assuming again a uniform density of 400 kg m⁻³, consistent with the spectral data that indicates the smooth patch shares a similar composition with the surrounding terrain, this corresponds to a mass of (15.4 ± 4.9) × $10^{10}$ kg for the cliff region. This estimate corresponds to a hypothetical end-member case in which the cliff-bounded recess did not exist and the surface formed a uniformly layer up to 50 m in height. It should not be interpreted as a literal reconstruction of the pre-existing topography, but rather as an idealized geometric configuration used to constrain the maximum volume potentially involved. The actual evolutionary pathway was likely more complex and may have involved localized failure, heterogeneities, or pre-existing relief. The estimate is therefore intended solely as a reference for assessing the scale of material loss or redistribution in this area.

A summary of these direct and inferred measurements is presented in Table 1. Caution is warranted in interpreting the statistical uncertainties. For example, estimating thickness involves errors related to the emission and incidence angles inherent to the camera's pointing accuracy (difficult to quantify), as well as subjective decisions about the boundary of diffuse shadows. Additional errors arise from projecting images onto a three-dimensional model, which itself contains intrinsic inaccuracies. For these reasons, we emphasize that these measurements should be regarded as order-of-magnitude estimates rather than precise values.

**Table 1. Estimated values for the area, thickness, and volume of the smooth patch and the cliff region. The mass is obtained using a density of 400 kg/m$^3$. Reported uncertainties include boundary-placement uncertainty, repeatability (expressed as the standard deviation of independent measurements), and formal error propagation. However, additional systematic effects are not fully quantified so the values should be regarded as conservative, first-order estimates.**

| | **Large smooth patch** | **Cliff region** |
|---|---|---|
| **Thickness (km)** | 0.025 ± 0.005 | 0.054 ± 0.017 |
| **Area (km$^2$)** | 1.720 ± 0.010 | 7.140 ± 0.010 |
| **Volume (km$^3$)** | 0.043 ± 0.009 | 0.386 ± 0.121 |
| **Mass (kg x $10^{10}$)** | 1.72 ± 0.34 | 15.4 ± 4.9 |

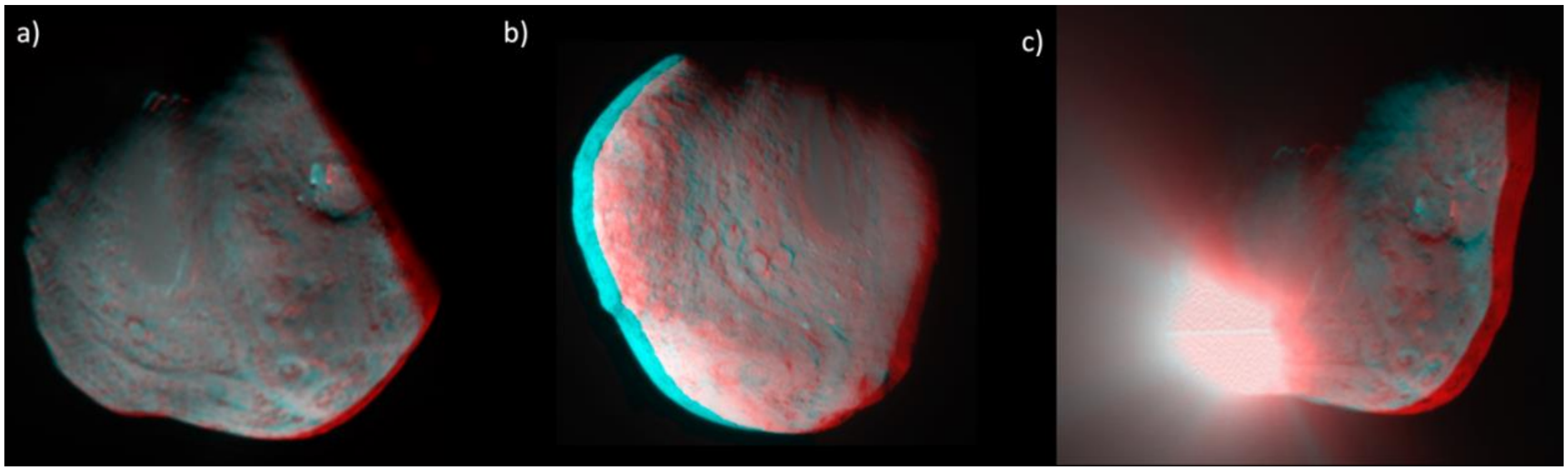

**Fig. 7. Anaglyphs to visualize the three-dimensional structure of the large smooth patch. These are composed of: a) cyan ITS image 5070405_9000639 paired with the red deconvolved HRI image 5070405_9000904, b) cyan image 30034 combined with red image 30035 from SDN, and c) cyan post-impact 5070405_9000999 overlaid with red image 5070405_9001012, both MRI images. The resultant three-dimensional perspective distinctly illustrates that the central region of the smooth patch dips lower than its surrounding sides, forming a discernible U-shape.**

### 3.3 A thorough comparison between Deep Impact and Stardust-NExT images

Because DI and SDN acquired images separated by approximately one orbit in time, we can study ice sublimation and analyze its evolution on 9P/Tempel 1. The inspection of both sets of images confirms that there are differences. However, due to differing viewing geometries and camera characteristics such as dynamic range, thermal stability, or spatial resolution, it is not straightforward to confirm whether these differences are real.

Our approach involves projecting the images onto the shape model and blinking them to identify apparent changes. First, we document all observed differences. Then, we assess whether these differences are real or false positives based on their location, morphology, and consistency across multiple images. Only the most evident and reliable real changes are presented in this section, while ambiguous or non-obvious variations are excluded. As an illustrative example, Fig. 8 displays the projected 30036 SDN and 5070405_9000673 DI images. Additionally, a blink animation of the images shown in Fig. 8a–b, as well as the labeled version corresponding to Fig. 8c–d, is provided in Appendix Fig. A4.

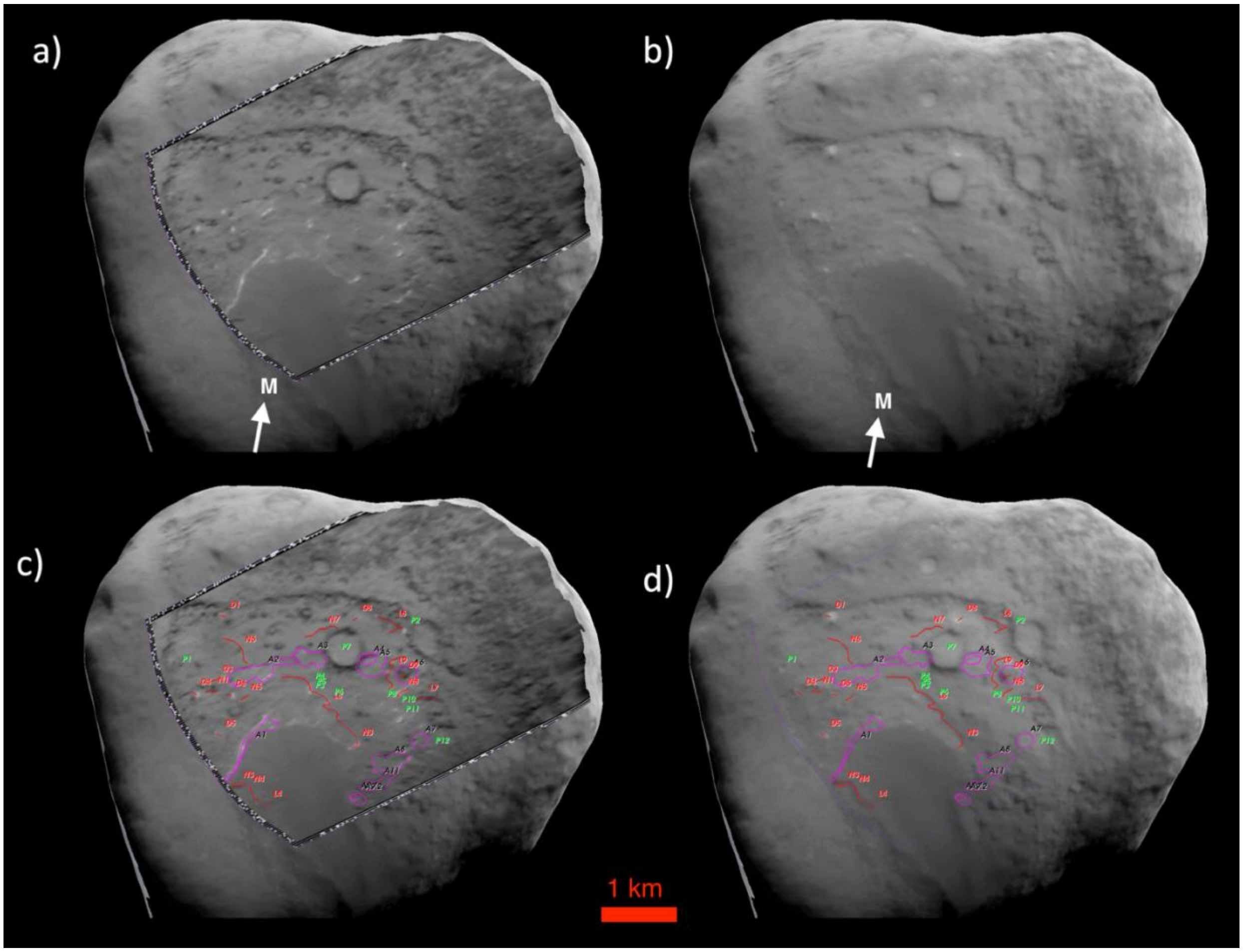

**Fig. 8. a) 30036 SDN NavCam image projected onto the 9P/Tempel 1 shape model. b) Same as (a) but with 5070405_9000673 DI ITS image overlaid. Panels (c) and (d) correspond to the same views shown (a) and (b) respectively, but display the labels referenced in the main text. The white arrows indicate the meridian (see Fig. 2 for reference). A blink animation is provided in Appendix Fig. A4.**

We labeled pairs of bright spots that are present in both DI and SDN images but show an offset as D1-8. We hypothesize that these bright spots are reflections from ice, with the offsets resulting from changes in incidence and emission angles caused by surface morphology alterations after sublimation.

Additionally, we labeled another group of bright spots that appear only in the DI image but are absent in the SDN image as P1–12. If these are reflections, it is possible that after ice sublimation, the new incidence and emission angles rendered them invisible from the SDN camera's viewpoint. Linear structures that are visible in the DI image but not in the SDN image are labeled as N1–9, while edge offsets are marked as L1–9.

Finally, we identified regions where sublimation has reduced parts of the large smooth patch, labeling them as A1–8. The most prominent of these is A1, a portion of the smooth patch edge that has completely sublimated, occupying an area of approximately 0.04 $km^2$. Since the thickness of this feature is ~25 m, we can establish a sublimation rate of at least $10^6$ $m^3$ per orbit –an order of magnitude higher than the estimate by Thomas et al., (2013) who assumed flow thickness between 8 and 15 m. However, given that the methodological details underlying their thickness assumption were not explicitly described, the origin of the discrepancy remains unclear.

### 3.5 Modeling solar irradiation

As described in Section 2.2, we computed the incoming solar radiation received by 9P/Tempel 1 over a full orbital period. The results of this calculation are shown in Fig. 9, alongside DI and SDN images for comparison, all rendered from the same viewing geometry. Note that in this computation, we also account for regimes in which the incident solar irradiation is insufficient for sublimation to be the dominant process. Thermal cycling, subsurface heat propagation, and diffusive gas transport through porous layers can progressively modify the internal structure, volatile distribution, and surface morphology of cometary nuclei over long timescales (Rickman et al. 1990, Prialnik 1992, Enzian et al. 1997). Therefore, the absence of dominant sublimation does not imply negligible long-term evolution.

*Large Smooth Patch*: This region is clearly visible in both the DI image (Fig. 9a) and the SDN image (Fig. 9b), and is identified as the area receiving the lowest insolation across the entire surface.

*Cliff Region, Northern Smooth Unit, and Terraces*: Insolation in this region is approximately three times higher than that received by the large smooth patch (Figs. 9a, 9c, and 9d).

*Equatorial Smooth Unit*: This is the most strongly irradiated area due to its location near the equator. Unfortunately, the absence of additional DI or SDN images from other perspectives limits a more complete characterization of this feature.

We emphasize that the simulation assumes 9P/Tempel 1's current orbital configuration to be broadly representative of its recent dynamical history. As such, it should be regarded as a first-order illustrative framework rather than a detailed thermophysical reconstruction.

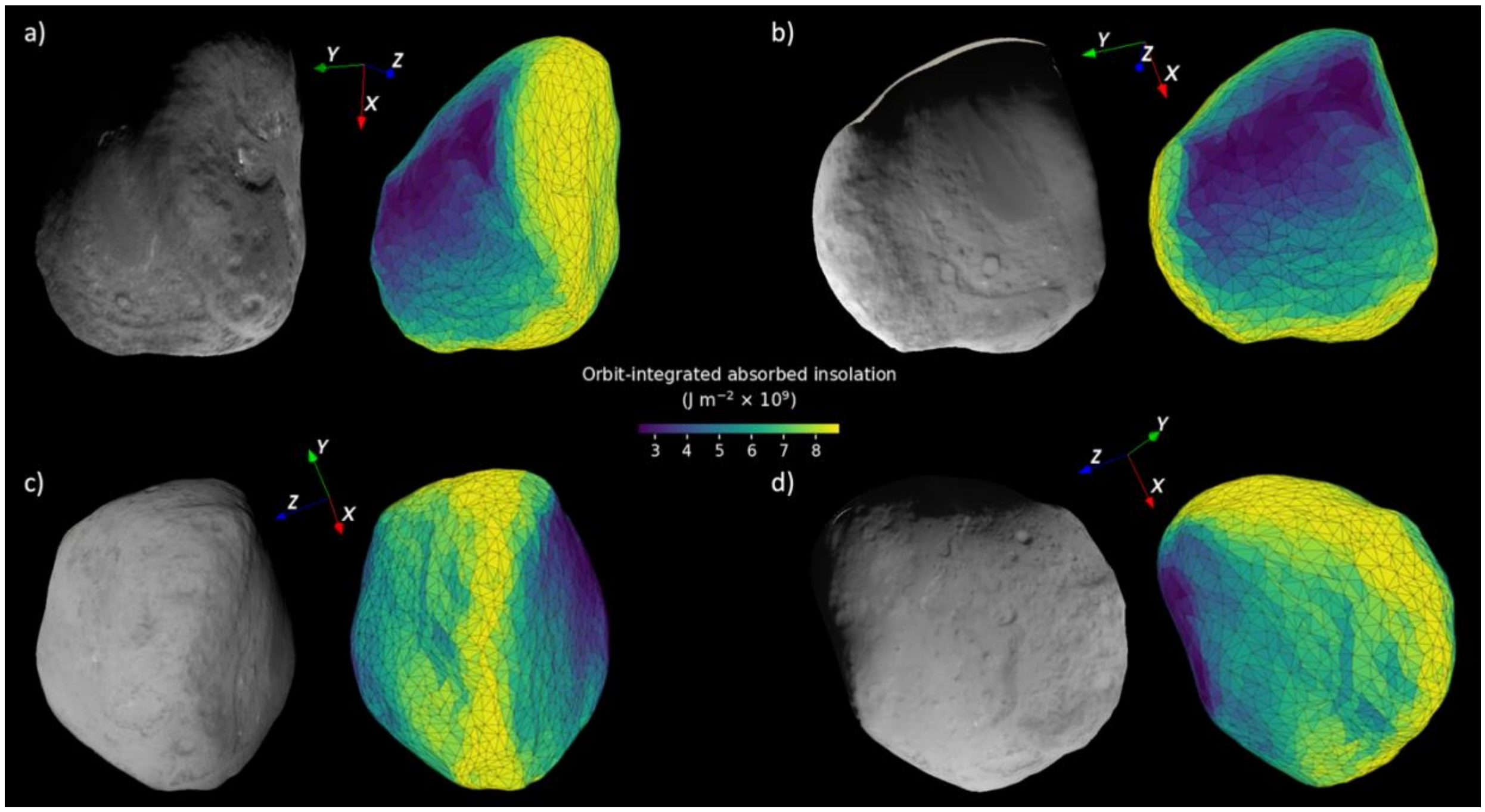

**Fig. 9. Total cumulative solar insolation (J/m$^2$) after a complete orbit of 9P/Tempel 1. An observed image is placed next to each figure to aid in interpretation: (a) corresponds to deconvolved 9000909 HRI, while (b), (c), and (d) are 30036, 30037, and 30039 NavCam images, respectively.**

## 3.4 Gravitational flow simulation

As explained in Section 2.2, we explored the gravitational flow hypothesis using the updated shape model and a kinematic flow simulation. To systematically assess all potential source regions, the surface of 9P/Tempel 1 was discretized using a 200 m grid, and each grid node was treated as an independent initial condition (“seed”). For each seed location, the flow was propagated iteratively to adjacent facets of lower gravitational potential, following the local potential gradient, until no further descent was possible. Each run therefore represents one kinematic solution under the assumed gravitational field.

All solutions initiated in the southern hemisphere converge toward the gravitational potential well corresponding to the location of the large smooth patch, indicating that this region constitutes a global topographic low within the modeled potential field.

The most intriguing results occur when the seed is placed in the equatorial region near the antemeridian, between longitudes 160° and 180° W (see Fig. 10, panel #1). In this configuration, the simulated flow not only fills the area corresponding to the large smooth patch, but also spreads across multiple faces of 9P/Tempel 1. Fig. 10 shows three stages of the simulation, in which the flow is represented in grey: panel #1 corresponds to the initial state with the seed already placed (hereafter referred to as the “seed region”); panel #2 illustrates an intermediate stage; and panel #3 presents the final outcome. In this example, the seed is restricted to latitudes between −30° and +20°, but other close latitude ranges yield comparable results, as long as they remain near the this equatorial region. The key parameter is longitude. A step-by-step animation of the simulation, together with a rotated view of the final result, is presented in Appendix Fig. A5. The rotated animation progressively changes the observer's perspective to visualize the flow from multiple viewing angles, covering all sides of the nucleus rather than only the southern hemisphere.

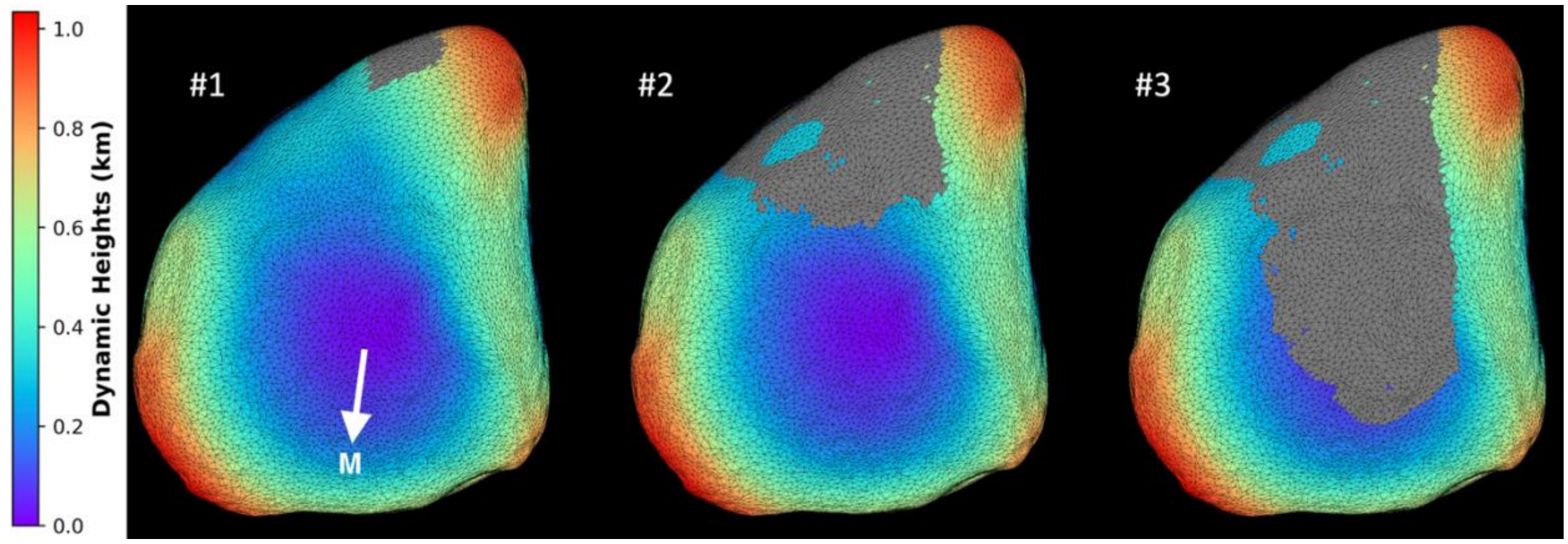

**Fig. 10. Simulation of flow evolution over the surface of 9P/Tempel 1. Panel 1 shows the initial state, with the seed region placed between 160° and 180° W in longitude and −30° and +20° in latitude. Panel 2 corresponds to an intermediate stage of the simulation, while 3 displays the final outcome. The blue-to-red color scale on the facets indicates the gravitational potential, expressed in terms of dynamic heights. Gray facets represent the simulated flow region. To optimize computation time, the shape model was downsampled by a factor of 100, resulting in a mesh of 31,456 facets. A latitude–longitude grid, displayed from the same viewing perspective, is provided in Fig. 2 for geographic orientation (the white arrow indicates the meridian). See Fig. A5 for the corresponding animation of this simulation.**

Importantly, the grey facets in Fig. 10 panel #3 represents the area that is gravitationally reachable by material moving downslope, rather than implying uniform deposition. Consistent with the observed morphology, most of the mobilized material should have accumulated preferentially in local gravitational minima.

In Fig. 11, we compare the simulation results with the observed images. Panel (a) shows how the equatorial smooth unit could be connected to the large smooth patch, assuming both are shaped by gravitational flow. Similarly, panel (d) illustrates the same interpretation applied to the northern smooth unit. Unfortunately, the seed region was not directly imaged by either DI or SDN, and its morphology cannot be analyzed.

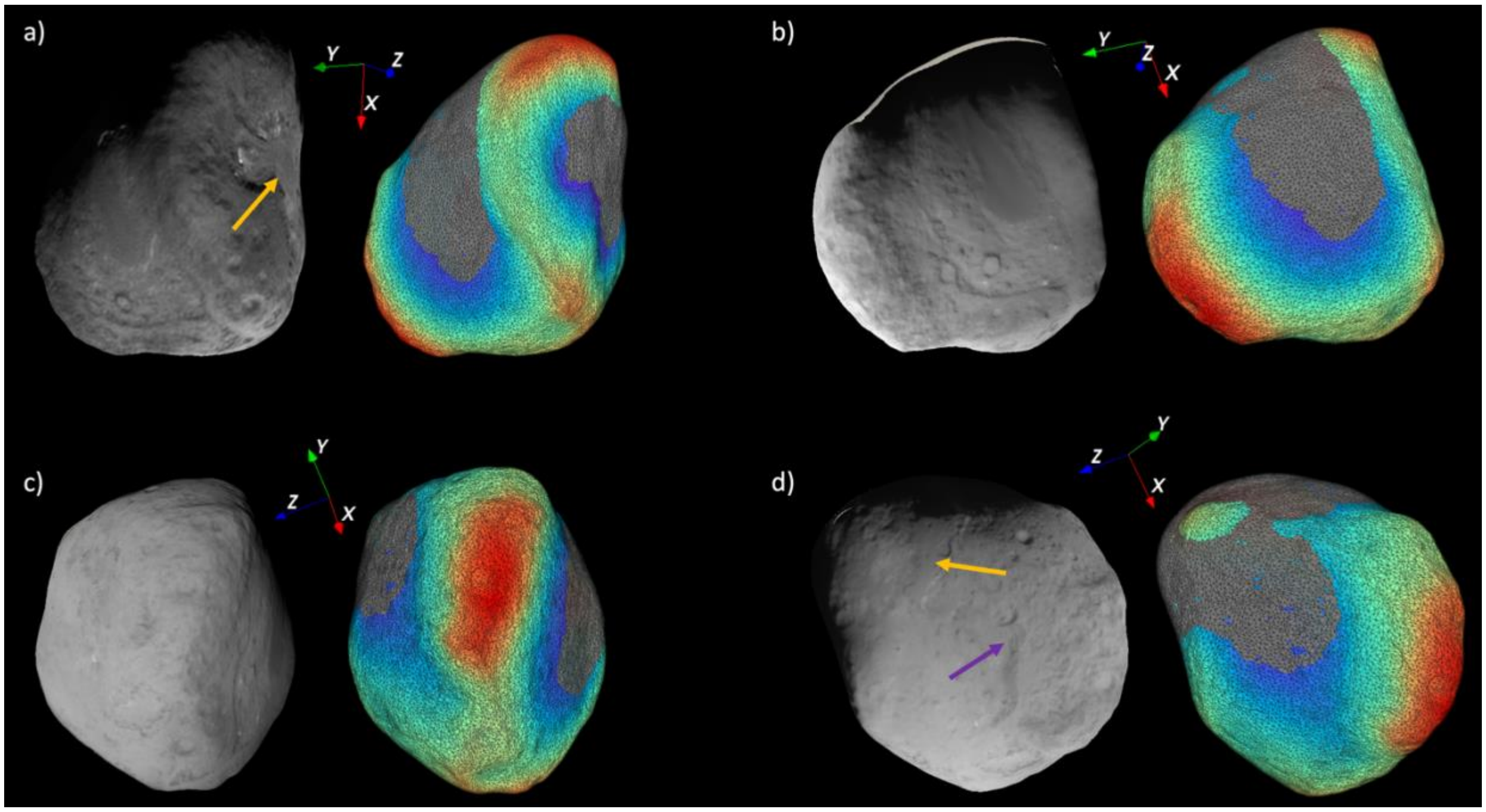

**Fig. 11. Same as Fig. 10, but with the shape model reoriented to match the viewing geometry of observed images for direct comparison (in all panels, the Z-axis indicates the north direction): panel (a) corresponds to the deconvolved 9000909 HRI image, where the orange arrow marks the location of the equatorial smooth unit; panels (b), (c), and (d) show NavCam images 30036, 30037, and 30039, respectively. In panel (d), the orange arrow points to the northern smooth unit, while the purple arrow indicates the terraces featuring up to five distinct steps. This simulation connects all the smooth units and the mass-wasting features on the opposite northern face as the outcome of a single event.**

## 4. Discussion

Our results support an endogenous origin for the large smooth patch, which appears compositionally similar to the rest of the surface. Topographic measurements indicate it is a ~25 m-thick deposit bounded by a ~50 m-high cliff, with a lobate U-shape morphology suggestive of gravitational flow.

Kinematic flow simulations further support this interpretation and point to a single event linking the large and secondary smooth patches with the topographic terrace features on the opposite face. Notably, given the rarity of such phenomena, invoking a one-time trigger is sufficient, without the need to postulate repeated exceptional conditions. According to this scenario, the cliff may have formed as a consequence of material redistribution associated with the flow and later became exposed as sublimation progressively removed part of the deposited material.

In this context, the terraces observed near the northern smooth unit could be consistent with subsequent mass-wasting processes. If material was transported downslope and accumulated along a margin, the later retreat of the deposit due to sublimation could have destabilized the resulting scarp. Such destabilization may have contributed to episodic collapses, producing the present step-like terraces. However, other processes including activity-driven slope modification, may also have contributed to the current morphology (Farnham et al. 2013, Steckloff & Samarasinha 2018).

Insolation patterns support the above hypothesis: the low insolation received by the large smooth patch is consistent with its preservation, allowing the deposited material to remain relatively intact. In contrast, the northern smooth unit, the adjacent cliff, and the terraced region receive significantly higher solar radiation, which is compatible with the idea that these features were subject to enhanced sublimation over multiple orbits.

It would be particularly interesting to further investigate the equatorial smooth unit. As this is the most irradiated region on the surface, one might expect to observe another cliff or residual structure from a preexisting flow as in the others. However, the lack of high-resolution images from varied viewing geometries prevents a proper analysis. Additionally, this unit lies adjacent to a crater- or pit-like feature, in the region where exposed water ice was previously detected (Sunshine et al., 2006). It is possible that this other feature played an important role in the current surface configuration. Nonetheless, we have little data to support more definitive interpretations.

Based on this evidence, we propose that the smooth patch may have resulted from a gravitational flow originating in the equatorial region near the antimeridian. The material then migrated toward the meridian, potentially modifying or reworking the pre-existing surface along its path. Subsequent sublimation may have progressively exposed a depressed area associated with this redistribution, as well as a steep scarp at the leading edge of the former flow. However, this scenario raises two key questions: What mechanism triggered this flow, and when did it occur?

### 4.1 Temporal Estimate

We measured a minimum sublimation rate of 0.04 km² per orbit and estimated that the cliff region spans approximately 7.14 km². Assuming this entire area was originally covered by a smooth flow and that sublimation has proceeded at a constant rate, we estimate that the material would have taken about 178 orbits to sublimate, yielding roughly 1,000 years.

To complement this analysis, we provide an independent estimate based on mass loss considerations. According to Feaga et al., (2007), the water production rate of 9P/Tempel 1 during the Deep Impact flyby was $4.6 \times 10^{27}$ molecules/s. Assuming that the nucleus is composed entirely of water ice, we apply the method described in Section 2.2 to compute the accumulated solar irradiance, but in this case restricting the calculation to heliocentric distances smaller than 3 AU. We adopt this cutoff as a practical threshold, acknowledging that water-ice sublimation may still occur beyond ~3 AU (Meech & Svoren, 2004; Prialnik et al., 2004), but is unlikely to dominate the overall activity at larger heliocentric distances. Under this condition, we find that the cliff-bounded region receives approximately 2.9% of the total orbit-integrated solar energy. Based on this fraction, we estimate that complete sublimation of the total mass of the cliff region would require on the order of ~1,200 years.

This solar irradiance fraction, however, implicitly assumes that the orbital parameters of the comet have remained unchanged over the past several thousand years. If instead we adopt a simpler geometric scaling, noting that the cliff region represents about 6% of the total surface area of the nucleus (~119 km²; Thomas et al., 2007), its relative contribution to the total water production would also be ~6%, yielding a shorter sublimation timescale of roughly 600 years.

In either case, these estimates should be regarded as order-of-magnitude approximations. They do not account for several important factors, including the likelihood that part of the flow consisted of dust rather than pure ice, reduced sublimation efficiency due to porosity, and the contribution of other volatile species such as CO, HCN, $CH_3OH$, CS, or $H_2S$ to the overall mass loss (Crovisier et al., 2009).

Taken together, both estimates converge on a formation timescale between 600 and 1,200 years ago, indicating that this is a relatively recent event in the comet's history. For comparison, Thomas et al., (2013) based on scarp geometry and the erosion of specific topographic features, yielded a minimum age of ~150 years.

### 4.2 Causal Estimate

Although the available data constrains are insufficient to identify a preferred mechanism, below we outline several plausible scenarios, which are also summarized in Table 2:

The most studied comet at present is 67P/Churyumov-Gerasimenko. For this comet, there is evidence of a **seasonal fallback process** (Keller et al., 2017, Jindal et al., 2022) that explains the smooth terraces found in the northern hemisphere (Birch et al., 2017). The obliquity and eccentricity of 67P/Churyumov-Gerasimenko's orbit result in more particles being released from the nucleus during the southern summer (Thomas et al., 2015). These particles follow ballistic trajectories, with a significant fraction of centimeter-sized or larger particles eventually falling back to the surface. They accumulate in the cold, gravitational lows of the northern (winter) hemisphere, forming the smooth terrains that evolve on short timescales. Groussin et al., 2015, 2019; Jindal et al., 2022 found that the Imhotep region presents a triple-scarp system, showing very evident changes over short timescales (weeks and months).

To evaluate whether this process could explain the formation of the smooth terrains on 9P/Tempel 1, we applied the dust fallback simulation developed for comet 67P/Churyumov–Gerasimenko by Kloos et al., (2025). The results are shown in Fig. 12. While there is net particle accumulation on both the southern hemisphere—where the large smooth patch is located—and the northern hemisphere—where the northern smooth patch lies—this accumulation does not correlate locally with the smooth terrains (e.g., Fig. 12b). In the case of the equatorial smooth patch, we observe neither net particle accumulation nor any morphological correlation with the modeled dust exchange. Deposition processes are generally expected to produce smooth surfaces rather than steep topographic relief. Therefore, it remains unclear how a purely deposition-driven scenario could account for the formation of the observed cliff, which likely requires either removal of material or structural collapse. It is worth noting, however, that unlike for 67P/Churyumov–Gerasimenko, we lack continuous temporal imaging of 9P/Tempel 1, and the comet's current spin axis may not reflect its past orientation.

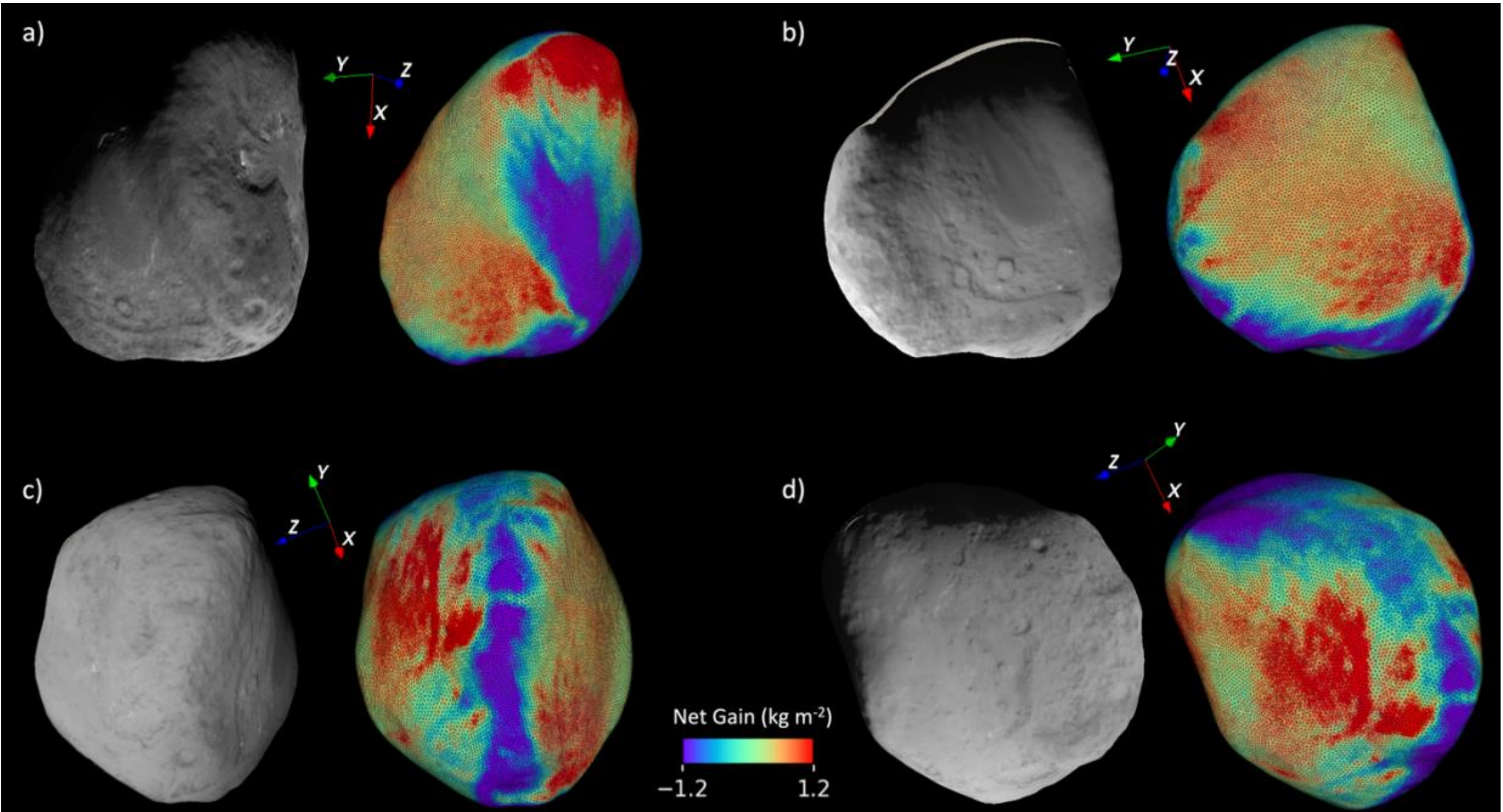

**Fig. 12. Net accumulation of dust material across one orbital period. An image is placed next to each figure to aid in interpretation: (a) corresponds to deconvolved 9000909 HRI, while (b), (c), and (d) are 30036, 30037, and 30039 NavCam images, respectively.**

**Cryovolcanism**, which involves the mobilization of fluid phases of water or other volatiles that subsequently freeze solid on the surface, is the simplest and most direct explanation for the observed flow. However, cryovolcanism requires a source of energy to increase internal pressure and/or melt subsurface ice. This phenomenon has already been suggested for objects with large heliocentric distances that are similar to comets (but of larger size), such as Charon or Quaoar (Cook et al., 2007; Jewitt & Luu 2004). However, the mechanisms driving these events are still not well understood.

In general, for large bodies, cryovolcanic activity is typically attributed to **either tidal** or **radiogenic heating**. In the case of comets, however, the radioactive decay of short-lived isotopes is considered inefficient. For example, Choi et al., (2002) demonstrated that this decay, such as from $^{26}$Al, can only raise temperatures to about -93°C, and even then, it remains localized in the nucleus. Moreover, it would only have been relevant during the early accretion phase of the nucleus and is not expected to contribute to its present-day thermal evolution. On the other hand, the tidal heating mechanism in larger bodies results from synchronous rotation and scales with the satellite's radius, so this process does not readily apply to the present case.

Cometary ice is thought to have formed predominantly as amorphous water ice in the early Solar System rather than crystalline (van de Bult et al., 1985; Whittet 1992; Prialnik & Jewitt 2022). Amorphous ice is a metastable form created when water vapor condenses at extremely low temperatures. In this state, the water molecules are unable to rearrange into more energetically stable positions due to insufficient energy or mobility, resulting in a highly disordered solid. If an event causes the molecules to reorganize, they will transition into lower-energy configurations, releasing heat in an **exothermic reaction**. This hypothesis offers a potential internal heat source for cometary cryovolcanism, but it also raises important questions, such as why similar flows have not been identified on other comets, what fraction of amorphous ice may actually be present today, if any, and which mechanisms could realistically trigger the phase transition. Moreover, laboratory experiments by Kouchi & Sirono (2001) demonstrated that even modest volatile impurities ($\gtrsim 2\%$ molar fraction of species such as $CO_2$) can substantially reduce the effective exothermicity of crystallization and may even render the process net endothermic.

A process that may be relevant in this context is the fluidization model proposed by Belton & Melosh (2009), which relies on the amorphous-to-crystalline phase transition of water ice, whereby cometary material becomes fluidized by the release of CO and/or $CO_2$ gas from depths of approximately 30–100 m. According to this model, such events occur intermittently, and the authors estimate that the resulting smooth terrains should be younger than 700 orbits (<4000 years). However, in addition to the limitations discussed above regarding the fraction of amorphous ice and the potential reduction in exothermicity due to the presence of volatile impurities, this mechanism presents further challenges. In particular, it predicts that such events should occur repeatedly and at random locations across the nucleus, and therefore does not naturally account for the apparent spatial connection between the different smooth features. Moreover, it relies on specific physical parameters such us the depth of the amorphous/crystalline boundary, a substantial fraction of $CO/CO_2$ in the interior, or an adequate internal gas pressure, that are currently poorly constrained and lack direct observational validation.

Another possibility involves a **pressurized phase-change** in an internal layer of 9P/Tempel 1. On the surface, the absence of atmospheric pressure makes the existence of a liquid phase impossible, but what if non-solid phases of cometary material had existed temporarily in the past? The melting of inner ice, or simply an increase in temperature and/or pressure sufficient to change viscosity or friction coefficient, could cause displacement and drag of materials, which in turn, due to heating by friction, could acquire sufficient viscosity to be reaccumulated in a semi-fluid state. De Almeida et al., (2016), in their studies of observed outbursts in comet 17P/Holmes, proposed that after an initial heating phase and the sublimation of water ice to form water vapor, some of this vapor could refreeze before escaping. This would block pore space and increase the comet's tensile strength in or near the outer crust region, preventing the release of newly liberated subsurface volatile gases, which would ultimately be discharged in explosive episodes. Such a model could account for the sudden explosive outburst activity observed before and after perihelion. They estimate that internal gas pressures on the order of ~1 kPa are necessary to rupture a locally strengthened crust, while Reach et al. (2010) inferred tensile strengths between 10 kPa and 200 kPa on scales exceeding 10 m for comet 17P/Holmes. Note that these values correspond to locally sealed regions, rather than to the bulk mechanical strength of cometary nuclei, which is generally inferred to be much lower. If something like this occurred, the temperatures and pressures generated readily surpass the triple point requirements, and it could have allowed the upper ice mass to move in the direction of the gravitational gradient. However, we must acknowledge that, to date, there is no evidence beyond speculation to support these ideas. In fact, there is not even evidence of such changes following the DI mission's impact on 9P/Tempel 1.

We can also consider the possibility of an **impact-triggered** event. However, cometary parent bodies are believed to have formed in low-density regions during the early stages of the Solar System, leading to significantly fewer collisions and, consequently, much lower impact energies for heating (Davidsson et al., 2016). Moreover, numerical and experimental studies indicate that even catastrophic impacts produce only highly localized heating, with negligible thermal alteration at nucleus scale (e.g., Schwartz et al., 2018). The DI experiment itself further demonstrated that an artificial impact did not trigger sustained activity or large-scale structural modification of the nucleus (Sunshine et al., 2016). Therefore, although impacts cannot be completely ruled out, they are unlikely to provide sufficient internal heating or mechanical disruption to explain the formation of the large smooth patch. Additionally, given the inferred young age of this feature, the probability of such an event being responsible remains low.

As previously noted, tidal heating mechanisms operating in large synchronously rotating bodies are not applicable in this case. However, **gravitational perturbations** associated with close encounters with Jupiter may still affect small bodies such as 9P/Tempel 1. As demonstrated by comet D/Shoemaker–Levy 9 (Asphaug & Benz 1994), tidal stresses during close approaches can be sufficient to even disrupt cometary nuclei. In this case, although tidal heating is not expected to play a significant role, such orbital perturbations may induce internal stresses and, more importantly, abrupt changes in the comet's perihelion distance, which could act as potential triggering mechanism.

According to Yeomans et al., (2005), comet 9P/Tempel 1 has undergone dramatic orbital changes due to close approaches to Jupiter over the years. These changes include a perihelion shift from approximately 1.5 to 3.5 au and inclinations ranging from about 3 to over 15 degrees. Similarly, Ip et al., (2016) performed a backward orbital integration for 9P/Tempel 1, finding that abrupt changes in its orbit, caused by encounters with Jupiter, resulted in perihelion distances ranging from 4 to 1.4 AU. Fig. 13 shows the time evolution of the perihelion distance over the past 3,000 years (data kindly shared by Wing-Huen Ip and Yu-Chi Cheng through private communication). These data result from a backward orbital integration performed using the Mercury numerical package (Chambers 1999). The abrupt changes in perihelion observed are caused by close encounters with Jupiter. The shaded region in the figure corresponds to the time window that we proposed as the formation period of the large smooth patch (600 - 1,200 years ago). Remarkably, repeated perihelion variations (from ~4 to 2.3 au before returning to ~4 au again) align well with the timescale inferred in this study.

This behavior is consistent with the "semimajor-axis jumps" described by Lilly et al. (2024), who show that abrupt orbital reshaping events can significantly enhance the average per-orbit heating and potentially initiate cometary activity. In this context, recurrent orbital perturbations may serve as an impulsive trigger for mechanical destabilization and subsequent mass mobilization, possibly acting in concert with one of the candidate mechanisms presented above. Jupiter's tidal forces, however, are several orders of magnitude weaker than 9P/Tempel 1's surface gravity and therefore would not govern the flow trajectory itself. Notably, orbital integrations by Ip et al. (2016) indicate that 9P/Tempel 1 is unique among the Jupiter-family comets analyzed (67P/Churyumov–Gerasimenko, 81P/Wild 2, 103P/Hartley 2, and 19P/Borrelly) in exhibiting such pronounced perihelion excursions (see their Fig. 9). Although not conclusive and insufficient on its own to resolve the internal physical mechanism, this scenario provides a plausible dynamical trigger within the broader framework discussed here.

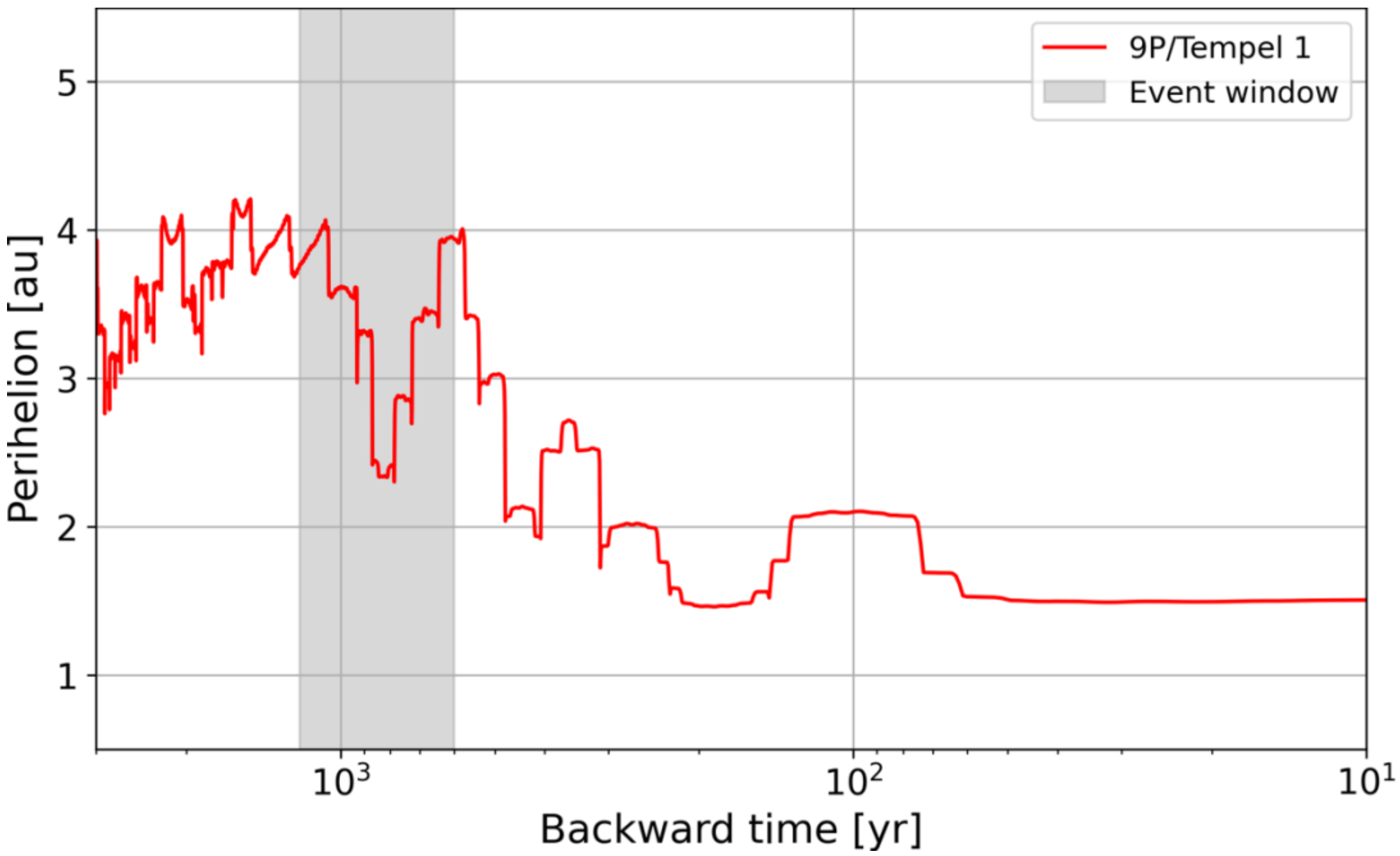

**Fig. 13. Data from Ip et al. (2016) showing the past orbital evolution of 9P/Tempel 1 over the last 3,000 years. The shaded area marks the interval between 1,200 and 600 years ago, corresponding to the estimated formation period of the large smooth patch. Notably, this timeframe coincides with the most abrupt orbital changes, during which the perihelion decreased from ~4 to 2.3 au before returning to ~4 au.**

It is interesting to note that the region where the flow must have emerged on 9P/Tempel 1 in order to connect the two smooth patches and the remnants of mass wasting as part of a single event (longitudes 160° - 180° W, see Fig. 10 panel #1) is a thin area located in the narrowest part of the equatorial region. Although it is highly speculative, it is worth noting that comet 67P/Churyumov-Gerasimenko is a contact binary comet with a narrow connection in a similar region. If we hypothesize that comet 9P/Tempel 1 was once part of a contact binary system, similar to 67P/Churyumov–Gerasimenko, it is plausible that the observed flow resulted from the exposure of an internal region following structural modification or partial disruption potentially associated with one of its past close encounters. However, dynamical studies indicate that such components would likely retain similar orbital characteristics over extended timescales (e.g. Hirabayashi et al., 2016) and there is currently no observational evidence that 9P/Tempel 1 was ever part of a dynamically separable pair.

**Table 2. Overview of the Possible Mechanisms Explaining the Surface Redistribution of Material on 9P/Tempel 1. For each proposed mechanism, the table highlights the supporting arguments as well as the opposing arguments or limitations that challenge the validity of these hypotheses.**

| Hypothesis | Supporting Arguments | Opposing Arguments |
|---|---|---|
| **Dust/Ice Fallback (accumulation)** | Similar process documented on 67P/Churyumov-Gerasimenko | No clear correlation between observed features and areas of high dust accumulation |
| **Radiogenic heating** | It could potentially trigger cryovolcanism through the melting of internal materials | Radioactive decay of short-lived isotopes is considered inefficient; would only have been relevant during the early accretion phase |
| **Synchronous tidal heating** | It could potentially trigger cryovolcanism by generating internal frictional heat; known to occur on larger icy bodies like Charon or Quaoar | It requires sustained gravitational interaction and synchronous rotation with a massive body |
| **Exothermic phase transition** | Amorphous-to-crystalline ice transition can be exothermic and may provide internal heating | No clear evidence of similar transitions in other comets; modest volatile impurities can significantly suppress the net heat release |
| **Pressurized Phase-Change** | Internal melting and gas overpressure could displace the material, generating flows | Too speculative; no evidence of temporary liquid phases |
| **Impact** | Explains sudden surface changes | Low likelihood of recent impacts; even catastrophic impacts produce only highly localized heating |
| **Gravitational perturbations** | Repeated close approaches to Jupiter align with estimated timeline; semimajor-axis jumps have been linked to the initiation of cometary activity; does not require exotic processes | It does not resolve the internal physical mechanism; scarcity of analogous cases to fully understand this phenomenon |

## 5. Conclusions

The morphological, spectral, and numerical analyses presented in this work provide a solid body of evidence that supports the interpretation that the large smooth patch on comet 9P/Tempel 1 is the result of a gravitational flow phenomenon. This structure, which exhibits a lobate U-shape morphology and is bounded by a 50-meter-high cliff, has a measured thickness of approximately 25 meters. Its lack of significant spectral differences compared to surrounding terrains reinforces the hypothesis of an endogenous origin. Gravitational flow simulations compellingly demonstrate how a single event, originating near the equatorial region, could have formed not only the large smooth patch but also the secondary smooth patches and the mass-wasting features observed on the northern face of the nucleus. Temporal estimates place the formation of the smooth patch between 600 and 1,200 years ago—remarkably consistent with a period of abrupt changes in the comet's perihelion distance, caused by multiple close encounters with Jupiter. The gravitational stresses and perihelion variations associated with such encounters may act as mechanical perturbations capable of triggering the event. Although not conclusive, and with the underlying mechanism still unresolved, these findings identify a direction that merits further study once more robust observational constraints are available.

The implications of these findings extend beyond the interpretation of a single surface feature. Our results suggest that external dynamical processes—particularly gravitational interactions with giant planets—may indirectly influence the large-scale geological evolution of cometary nuclei by acting as a potential trigger for events such as the one that gave rise the smooth units observed on comet 9P/Tempel 1. This challenges the traditional paradigm in which cometary evolution is driven almost exclusively by solar-induced sublimation. The uniqueness of the feature observed on Tempel 1, which we propose is linked to its orbital history, reinforces the idea that a comet's geological and dynamical evolutions are fundamentally interconnected. However, we are still far from fully understanding this phenomenon. The absence of continuous temporal images and the scarcity of analogous features on other comets limit our ability to fully validate the theory. Future space missions will be essential for advancing our understanding of cometary evolution, delivering comprehensive dataset that may shed light on the complex mechanisms driving their behavior and long-term transformation.

## Acknowledgments

This work was supported by NASA Discovery Data Analysis grant 80NSSC18K1433 (T. L. Farnham, PI). J.L. Rizos acknowledges financial support from the Severo Ochoa grant CEX2021-001131-S, funded by MCIN/AEI/10.13039/501100011033, and from grant PID2021-126365NB-C21. The authors thank Wing-Huen Ip and Yu-Chi Cheng for providing the backward orbital integration data of Tempel 1 used in this study; Lori Feaga for providing the calibrated DI spectral data; Peter Thomas for providing data for the gravitational simulations; Carolyn Ernst for sharing the updated stereophotoclinometry-based shape model; and Jean-Baptiste Vincent for valuable discussions.

# Appendix

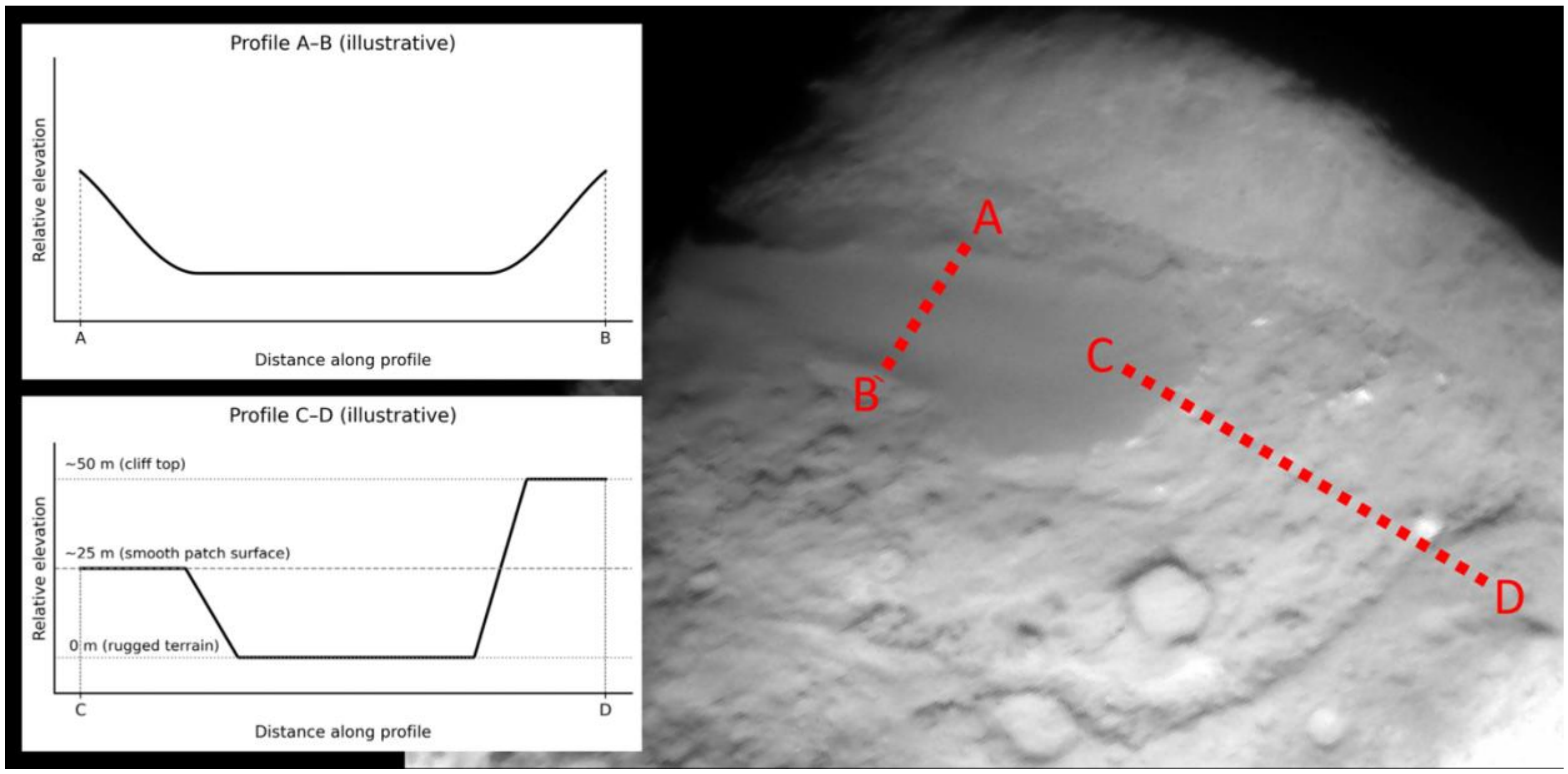

**A1. Stardust/NExT image 30035 centered on the large smooth patch on the nucleus of comet 9P/Tempel 1. The image shows the context of the smooth patch relative to the surrounding rough terrain and the adjacent cliff-bounded region. The dashed red segments indicate the locations of two representative transects, A–B and C–D. The left panels show illustrative cross-sectional profiles along A–B (top) and C–D (bottom), highlighting the concave morphology of the smooth patch, as well as the relative elevation of the surrounding cliff. The profiles are schematic and are intended solely to clarify the geometric relationships discussed in the text.**

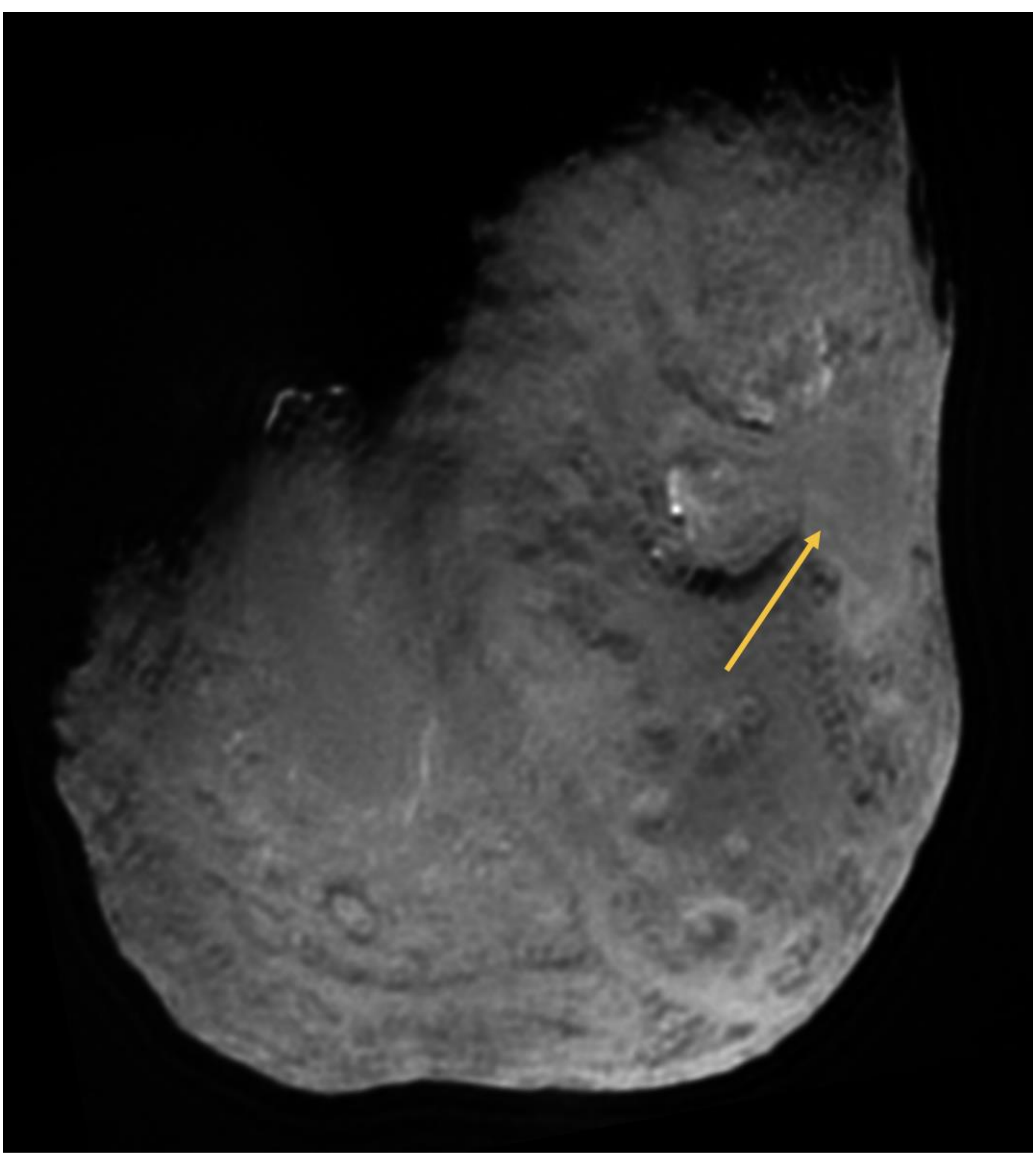

**A2. Deconvolved HRI image 9000909 showing the equatorial face of Tempel 1. The orange arrow marks the location of a secondary smooth unit, adjacent to the area where Sunshine et al. (2006) reported exposed water ice. In this work, we refer to this feature as the "equatorial smooth unit". The large smooth patch is also visible on the left.**

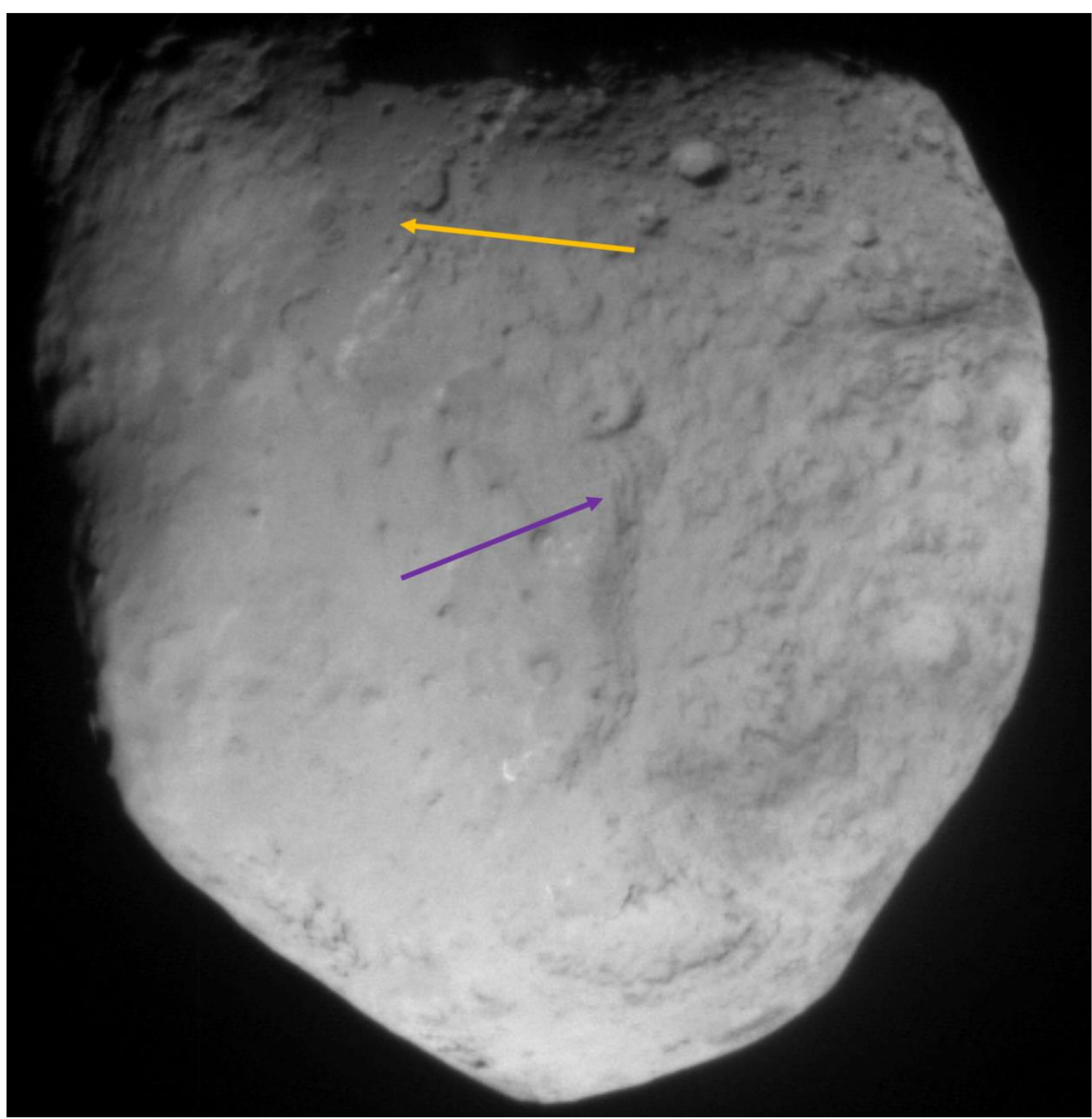

**A3. Stardust/NExT image 30039 showing the northern hemisphere of Tempel 1 (opposite to the side shown in Fig. A1). The orange arrow indicates the region where a smooth unit is visible. We refer to this feature as the “northern smooth unit”. The purple arrow points out the terraces featuring up to five steps.**

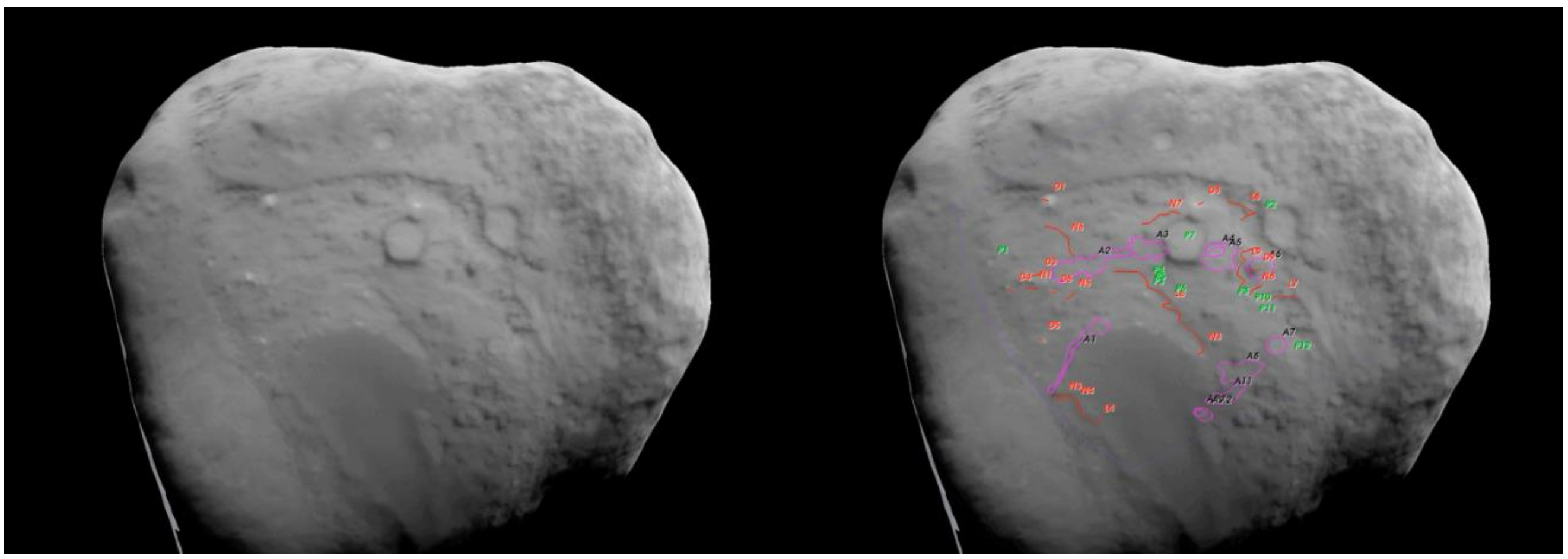

**A4. Left: Blink comparison between the SDN NavCam image 30036 and the DI ITS image 5070405_9000673, both projected onto the 9P/Tempel 1 shape model. It alternates between the two datasets to emphasize morphological changes between the DI and SDN epochs. Right: Annotated version of the blink comparison, used to identify and track surface modifications.**

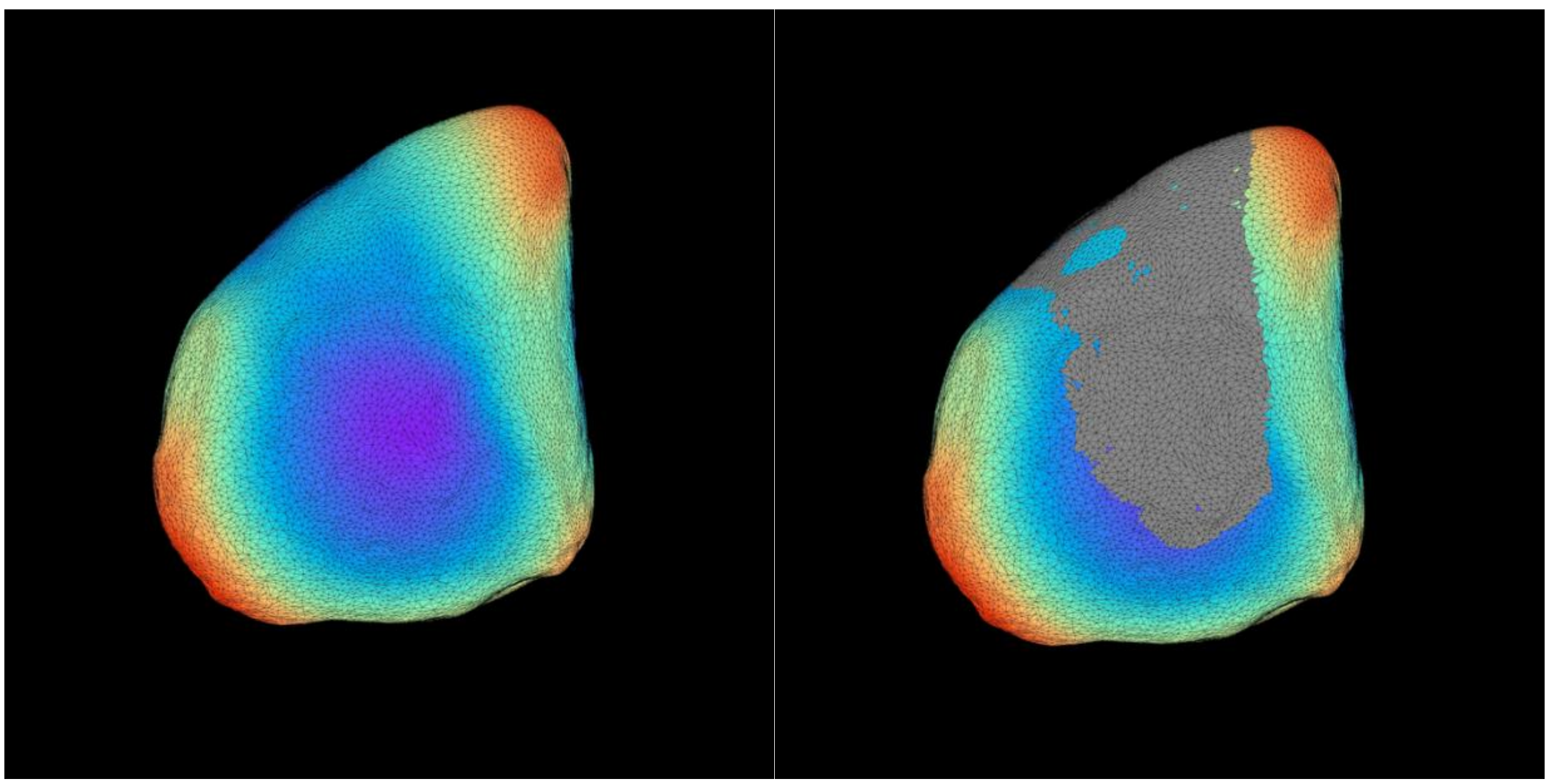

**A5. Left: Gravitational flow simulation focused on the southern hemisphere. The seed is placed in the equatorial region near the antimeridian (160°–180° W) and treated as the source of the flow. At each step, the flow propagates to adjacent facets with lower gravitational potential, and the process continues iteratively until no lower-potential neighbors remain. Right: Final extent of the simulated flow, shown from a rotated perspective to display its full spatial distribution across the nucleus. Colors represent gravitational potential, with cooler tones indicating lower values. The grey facets indicate those reached during the simulation.**